\documentclass[preprint,superscriptaddress,nofootinbib,preprintnumbers,amsmath,amssymb,11pt,aps,prd,longbibliography]{revtex4-1}
\usepackage[T1]{fontenc}

\usepackage[titles]{tocloft}
\cftsetindents{section}{0em}{2.5em}
\cftsetindents{subsection}{2.5em}{2.5em}

\usepackage[lofdepth,lotdepth,caption=false]{subfig}
\usepackage{amsfonts}
\usepackage{mathrsfs}
\usepackage{leftidx}
\usepackage{amssymb}
\usepackage{amsmath}
\usepackage{placeins}
\usepackage{relsize}
\usepackage{slashed}
\usepackage{multirow}
\usepackage[dvipsnames]{xcolor}
\usepackage{tikz}
\usepackage{array}
\usepackage{physics}
\usepackage{booktabs}
\usepackage{graphicx}
\usepackage{appendix}
\usepackage[caption=false]{subfig}

\usepackage{color}
\definecolor{SeaBlue}{rgb}{0.1,0.4,0.85}
\definecolor{DarkBlue}{rgb}{0.0,0.0,0.7}
\definecolor{NavyBlue}{rgb}{0.0,0.0,0.4}
\definecolor{Maroon}{rgb}{0.6,0.2,0.2}
\definecolor{SeaGreen}{rgb}{0.2,0.4,0.2}
\definecolor{Purple}{rgb}{0.7,0.1,0.6}
\definecolor{Red}{rgb}{0.8,0.2,0.2}
\definecolor{Black}{rgb}{0.0,0.0,0.0}
\definecolor{ZQS}{rgb}{0.01,0.227,0.451}
\definecolor{TS}{rgb}{0.588,0.31,0.05}
\usepackage[colorlinks,bookmarksnumbered, bookmarksopen, bookmarksopenlevel=3]{hyperref}
\hypersetup{
     colorlinks = true,
	linkcolor = NavyBlue,
    citecolor = red
}
\usepackage[T1]{fontenc}
\usepackage{titlesec}
\titleformat{\section}
  { \color{NavyBlue} \normalfont \scshape}
  {\thesection}{1em}{\bf \MakeUppercase}
\titleformat{\subsection} { \color{NavyBlue}  \normalfont\scshape}{\thesubsection}{1em}{\bf }{}
\titleformat{\subsubsection}
  { \color{NavyBlue}  \normalfont\itshape}{\thesubsubsection }{1em}{}{}

\usepackage{soul}
\usepackage{epsfig}
\usepackage{graphicx}               
\usepackage{url}
\usepackage{float}
\usepackage{animate}  

\newcommand{\be}{\begin{equation}}
\newcommand{\ee}{\end{equation}}
\newcommand{\bea}{\begin{eqnarray}}
\newcommand{\eea}{\end{eqnarray}}
\newcommand{\beq}{\begin{eqnarray}}
\newcommand{\eeq}{\end{eqnarray}}
\def\bit{\begin{itemize}}
\def\eit{\end{itemize}}
\def\ben{\begin{enumerate}}
\def\een{\end{enumerate}}

\newcommand\DN[1][\relax]{%
\ifx\relax#1\relax\else{}^{#1}\fi \!X}

\DeclareMathAlphabet\mathbfcal{OMS}{cmsy}{b}{n} 

\begin{document}

\count\footins = 1000 


\title{\texorpdfstring{\color{NavyBlue} \Large Theoretical bounds on dark Higgs mass in a self-interacting dark matter model with $U(1)'$}{Theoretical bounds on dark Higgs mass in a self-interacting dark matter model with U(1)'} }

\author{Song Li}
\email[]{lisong@itp.ac.cn}
\affiliation{ School of Physics, Henan Normal University, Xinxiang 453007, P. R. China}
\affiliation{Institute of Theoretical Physics, Chinese Academy of Sciences, Beijing 100190, P. R. China} 

\author{Jin Min Yang}
\email[]{jmyang@itp.ac.cn}
\affiliation{ School of Physics, Henan Normal University, Xinxiang 453007, P. R. China}
\affiliation{Institute of Theoretical Physics, Chinese Academy of Sciences, Beijing 100190, P. R. China} 

\author{Mengchao Zhang}
\email[Corresponding author, ]{mczhang@jnu.edu.cn}
\affiliation{Department of Physics and Siyuan Laboratory, Jinan University, Guangzhou 510632, P. R. China}

\author{Rui Zhu}
\email[Corresponding author, ]{zhurui@itp.ac.cn}
\affiliation{Institute of Theoretical Physics, Chinese Academy of Sciences, Beijing 100190, P. R. China} 
\affiliation{School of Physics, University of Chinese Academy of Sciences,  Beijing 100049, P. R. China}

\begin{abstract} 
Motivated by the null results of current dark matter searches and the small-scale problems, we study a dark sector charged by a spontaneous broken gauge $U(1)'$.  To explore the parameter space of this model, in addition to the consideration of the small-scale data, we also consider the theoretical bounds on the dark Higgs mass, with the upper bound coming from the tree-level perturbative unitarity and the lower bound from the one-loop Linde-Weinberg bound.   We deeply examine the dependence of the Linde-Weinberg bound on gauge choice and energy scale, and present a Linde-Weinberg bound that is gauge and scale independent. Combining the theoretical and observational constraints, we obtain the following ranges for the parameter space: the dark matter mass is 10-500 GeV,  the mediator (dark photon) mass is 0.5-5 MeV, the dark Higgs mass is 0.05-50 MeV, and the dark fine-structure constant is 0.001-0.4. We conclude that the dark Higgs in this model cannot be ignored in the phenomenological study of the dark sector.
\end{abstract}

\maketitle
\tableofcontents

\clearpage

\section{Introduction \label{sec:introduction} }

Numerous astronomical and cosmological observations have confirmed the existence of dark matter (DM)~\cite{Clowe:2006eq,Planck:2018vyg}, but so far we still know very little about its properties.
Weakly Interacting Massive Particles (WIMPs), which can naturally explain the DM relic density via thermal freeze-out and be embedded in new physics models like supersymmetry~\cite{Jungman:1995df}, have been searched for in  decades~\cite{Abercrombie:2015wmb,PandaX-II:2016vec,LUX:2016ggv,ATLAS:2017nga,ATLAS:2013ndy,CMS:2016gox,ATLAS:2017uis}. 
However, the null results of WIMP searches suggest us to consider more scenarios beyond the WIMP paradigm. 

A simple but useful way going beyond the WIMP scenarios is to enrich the DM sector by introducing a dark mediator particle. 
In such a so-called secluded DM scenario, generally there is no direct coupling between DM and visible particles, and the relic density is determined by the process of DM pair annihilation to mediator pair~\cite{Pospelov:2007mp}.  
Thus, unlike the WIMP scenarios where the relic density is directly related to the couplings between DM and visible particles, this scenario can easily avoid the strong constraints from the DM direct detection experiments. 

In a secluded DM scenario, a dark Abelian $U(1)'$ gauge interaction is usually introduced to stabilize DM and provide a vector mediator \footnote{The mediator can also be a light scalar in the singlet extensions of minimal supersymmetric model, see, e.g., \cite{Wang:2014kja,Wang:2022lxn,Wang:2022akn}.} called dark photon and labeled as $A'$.  
To escape limits from BBN or $N_{\text{eff}}$~\cite{Hufnagel:2018bjp,Ibe:2021fed,Depta:2020zbh,Ghosh:2023ilw}, dark photon generally need to be massive. 
In the literature, dark photon mass is often regarded as a free parameter and can be generated via the Stueckelberg mechanism. 
But it should be noted that Stueckelberg mechanism is actually the Higgs mechanism in Abelian gauge theory with the Higgs boson being decoupled. 
In the presence of massive dark photon, ignoring the dark Higgs may not be reasonable.

With a certain Higgs vacuum expectation value (VEV), the dark photon mass square and the Higgs boson mass square are proportional to the gauge coupling square ${g'}^2$ and the Higgs self-coupling $\lambda$, respectively. 
Therefore, only for $ \lambda \gg  {g'}^2$  can the dark Higgs be very heavy and thus be ignored in the dark sector.

However, the condition $ \lambda \gg  {g'}^2$ is generally unsatisfied in a reliable dark sector model if we require the theory to meet the common requirement of perturbative unitarity. 
Considering perturbative unitarity, $\lambda$ cannot be much larger than 1 and thus $\lambda \gg  {g'}^2$ can only happen for a very small ${g'}$. 
But, if the DM relic density is obtained through the freeze-out process, then $g'$ cannot be too small whether in the case of symmetric DM or asymmetric DM.
On the other hand, $g'$ can be quite small in a freeze-in scenario, but it would weaken the original motivation for introducing a dark mediator.

The assumption $ \lambda \gg  {g'}^2$ becomes even more untenable if we try to use the elastic scattering between the DM particles to solve the so-called small-scale problems.
Small-scale problems are a series of inconsistencies between the cold DM predictions and the observational data at small scales ( $\lesssim \mathcal{O}$(Mpc) )~\cite{Moore:1999gc,Moore:1994yx,Flores:1994gz,Oman:2015xda,Moore:1999nt,Klypin:1999uc,Boylan-Kolchin:2011lmk,Boylan-Kolchin:2011qkt}.
Previous studies have shown that these inconsistencies can be largely mitigated by a sizable and velocity-dependent elastic scattering cross section between the DM particles~\cite{Kochanek:2000pi,Andrade:2020lqq,Elbert:2016dbb,Fry:2015rta,Yoshida:2000bx,Moore:2000fp,Zavala:2012us,Elbert:2014bma,Rocha:2012jg,Peter:2012jh,Kaplinghat:2015aga,Tulin:2017ara,Vogelsberger:2014pda,Dooley:2016ajo,Dave:2000ar,Robertson:2018anx,Spergel:1999mh,Colin:2002nk,Vogelsberger:2012ku,Harvey:2015hha,Burkert:2000di,Sagunski:2020spe,Yoshida:2000uw}. 
Such kinds of models with large DM self-scattering cross-sections are also called self-interacting DM (SIDM) models in the literature. 
Relevant model building works can be found in ~\cite{Chen:2023rrl,Han:2023olf,Tulin:2013teo,Loeb:2010gj,Buckley:2009in,Foot:2014uba,Foot:2014osa,Bellazzini:2013foa,Feng:2009hw,Feng:2009mn,Wang:2016lvj,Ma:2017ucp,vandenAarssen:2012vpm,Kamada:2020buc,Ko:2014bka,Kitahara:2016zyb,Tulin:2012wi,Kainulainen:2015sva,Kamada:2018zxi,Duerr:2018mbd,Kamada:2018kmi,Aboubrahim:2020lnr,Ko:2014nha,Schutz:2014nka,Foot:2016wvj,Boddy:2014qxa,Bringmann:2013vra,Kamada:2019jch,Kamada:2019gpp,Kang:2015aqa,Tran:2023lzv,Borah:2021pet,Borah:2022ask,Borah:2021qmi,Borah:2023sal,Mahapatra:2023oyh,Borah:2024wos,Adhikary:2024btd,Chaffey:2021tmj} (for recent reviews, see, e.g.,  ~\cite{Adhikari:2022sbh,Tulin:2017ara}).  
In the scenario of dark sector with $U(1)'$ gauge interaction, the $U(1)'$ gauge boson $A'$ can naturally induce such a velocity-dependent cross section, provided that the mass of $A'$ ranges around $\mathcal{O}(1) - \mathcal{O}(10)$ MeV. 
But $g'$ also needs to be large enough to make a sizable cross section. 
Therefore, in the SIDM framework, as long as perturbative unitarity is considered, we cannot simply ignore the dark Higgs in the dark sector.

Another related interesting point is that the study of small-scale problems may help us to understand the nature of the dark sector, even if there is still no direct or indirect DM detection signals. 
To be more specific, the DM halos in different celestial systems (such as dwarf galaxies or galaxy clusters) can be regarded as different DM evolution systems. 
By fitting the observation data in these different celestial systems, it is hopeful to obtain information about the dark sector such as the masses of DM and mediator~\cite{Kaplinghat:2015aga,Correa:2020qam}. 
However, the mass of dark Higgs can not be revealed by this method, because the dark Higgs is not the mediator inducing the DM scattering inside a halo~\footnote{Precisely speaking, the dark Higgs can also be the mediator if DM is chiral instead of vector-like. But in this case there will be other issues like anomaly cancellation and thus the dark sector model cannot be so minimal.}. 
This situation is somewhat similar to the search for the Higgs boson in the Standard Model (SM), as the Higgs mass is the last parameter to be determined in the SM. 
During the search for the SM Higgs boson, we had already determined its mass range through theoretical considerations such as perturbative unitarity and vacuum stability before directly detecting the Higgs boson. 
These theoretical limits also provided motivations for the construction of the LHC. 
In this work, we aim to perform similar studies in the dark sector. 
By considering theoretical constraints in the dark sector, we will determine the mass range of the dark Higgs and highlight the significance of dark Higgs in the phenomenological study of dark sector. 


This paper is organized as follows.
In Sec.~\ref{sec:model}, we describe the DM model with $U(1)'$ gauge symmetry.  
In Sec.~\ref{sec:upper} and Sec.~\ref{sec:lower}, we examine the upper and lower bounds on the dark Higgs mass from perturbative unitarity and one-loop effective potential, respectively.  
In particular, in Sec.~\ref{sec:lower} we will discuss the dependency of gauge choice and energy scale in detail.
In Sec.~\ref{sec:fit}, the combined constraints on dark sector parameter space from observed data and theoretical requirements are presented.  
Finally we conclude this work in Sec.~\ref{sec:conclu}.

\section{A dark sector with spontaneously broken \texorpdfstring{$U(1)'$}{U(1)'}  \label{sec:model} }

The model we consider in this work is a dark sector charged under a gauge $U(1)'$, containing both dark matter and dark mediator: 
\begin{eqnarray}
\mathcal{L}_\text{Dark} = \bar{\chi}( i \slashed{D} - m_{\chi} ) \chi + (D_{\mu} S )^{\dagger} D^{\mu} S - \frac{1}{4} F'_{\mu\nu}  F'^{\mu\nu} - V_0(S)   \, ,
\label{model}
\end{eqnarray} 
with $\chi$ being the DM, $F'_{\mu\nu} = \partial_{\mu}A'_{\nu} - \partial_{\nu}A'_{\mu} $ being the field strength of dark photon and $S$ being the dark Higgs to make $A'_{\mu}$ massive. 
The covariant derivative is $D_{\mu} = \partial_{\mu} + i g' Q_i A'_{\mu}$. 
The $U(1)'$ charge of $\chi$ and $S$ can be different (labeled as $Q_{\chi}$ and $Q_{S}$).
For later convenience we can define a dark fine-structure constant $\alpha' \equiv {g'^2}/{4\pi}$. 
The tree-level potential $V_0$ is given by 
\begin{eqnarray}
V_0(S)  =  -\mu^2 S^{\dagger} S + \frac{\lambda}{4} (S^{\dagger} S )^2  \, . 
\end{eqnarray} 

After spontaneous symmetry breaking (SSB), $S$ gets a VEV, and then it can be re-expressed as
\begin{eqnarray}
S = \frac{1}{\sqrt{2}} \left( v_{\rm cl} + s + ia \right) ,
\end{eqnarray}
where $v_{\rm cl} \equiv 2\mu/\sqrt{\lambda}$ is the classical VEV of $S$ (at zero temperature),
$s$ and $a$ are the scalar and pseudo-scalar components of $S$, respectively.

To remove the unphysical components of gauge field, the gauge-fixing term and the accompanying ghost fields need to be added to the Lagrangian. 
In $R_{\xi}$ gauge, the gauge-fixing and ghost terms are given by
\begin{eqnarray}
\mathcal{L}_{\text{gf+gh}} = -\frac{1}{2\xi} (\partial_{\mu} A'^{\mu} - \xi Q_{S}g' v_{\rm cl} a )^2 - \bar{c} \left( -\partial_{\mu}\partial^{\mu} + \xi Q^2_{S} {g'}^2 v_{\rm cl}(v_{\rm cl}+s)  \right) c \, ,
\end{eqnarray}
where $c$ is the ghost field.


The mass squares of $s$ (dark Higgs), $a$ (Goldstone particle), $c$ (ghost field), and $A'$ (dark photon) after SSB and gauge-fixing are given by
\begin{eqnarray}
\label{mass_relation}
& & m^2_s = \frac{3}{4} \lambda v_{\rm cl}^2 - \mu^2 = \frac{1}{2} \lambda v_{\rm cl}^2  , \\
& & m^2_a = \frac{1}{4} \lambda v_{\rm cl}^2 - \mu^2 + \xi Q_{S}^2 g'^2 v_{\rm cl}^2 = \xi Q_{S}^2 g'^2 v_{\rm cl}^2  , \\
& & m^2_c = \xi Q_{S}^2 g'^2 v_{\rm cl}^2  , \  m^2_{A'} = Q_{S}^2 g'^2 v_{\rm cl}^2  \, .
\end{eqnarray}

For simplicity, we take $Q_\chi=Q_S=+1$~\footnote{Certainly, one can consider other values for $Q_\chi$ and $Q_S$. But the main conclusion in this work will not change much, provided the values of $Q_\chi$ and $Q_S$ are not too different. Generally speaking, the assumption $Q_\chi \approx Q_S$ is valid, because $Q_\chi \ll Q_S$ or $Q_\chi \gg Q_S$ are quite unnatural choices from a model building perspective.}.
Therefore, this dark sector model can be fully determined by 4 parameters: 
\begin{eqnarray}
m_\chi  \ , \ m_{A'} \ , \ g' \ \text{(or $\alpha'$)} \ , \ m_s
\end{eqnarray}
The task in the next two sections is to determine the allowed region of $m_s$ with certain value of $m_{A'}$ and $g'$.  
The DM mass $m_\chi$ will come into play in Sec.\ref{sec:fit} when we take small-scale data into account.

\section{Upper bound on dark Higgs mass from perturbative unitarity \label{sec:upper} }

In theories of symmetry breaking induced by scalar fields with non-zero vacuum expectation values, perturbative unitarity is an important tool for obtaining an upper bound of the corresponding Higgs boson mass. 
Although the violation of perturbative unitarity does not imply the failure of the theory, but as long as we believe that the theory can be treated perturbatively, then the corresponding Higgs boson mass can not take an arbitrarily large value. 
The most famous application of perturbative unitarity is the calculation of the upper bound of the Higgs boson mass in the SM by Lee, Quigg, and Thacker in 1977~\cite{Lee:1977yc,Lee:1977eg}. This mass bound guided numerous subsequent experiments searching for the Higgs boson, including the LHC experiment which was the first to successfully find the Higgs particle~\cite{ATLAS:2012yve,CMS:2012qbp}. 
In this section, we will provide a self-contained derivation for the commonly used perturbative unitarity bound, and then calculate the upper bound of the dark Higgs mass in our model. 
Unlike the usual derivation methods in the literature, our derivation does not based on the scattering theory in quantum mechanics (See~\cite{Jacob:1959at,devanathan1999angular,martin1970particle,Logan:2022uus} for relevant discussion).
The complete derivation with all the details is given in Appendix \ref{appA}.
Other relevant discussions about unitarity in dark sector can be found in ~\cite{Baek:2012se,Kamada:2022zwb}.

\subsection{A brief review of unitarity bound \label{subsec:UnitarityOnSubspace} }

For a normalized initial state $\ket{a}$ evolves to a normalized final state $\ket{n}$, the corresponding S-matrix element is $\mel{n}{S}{a}$.
Consider all possible final state $\ket{n}$, then due to the conservation of probability, we have 
\begin{eqnarray}
1=\sum_{n\in I}\abs{\mel{n}{S}{a}}^2=\sum_{n\in I}\mel*{a}{S^\dagger}{n}\mel{n}{S}{a}
\end{eqnarray}
where we use $I$ to label the set of all the possible final state. 
In the real case, we will only consider a subset of all possible final state (labeled as $I'$), and thus
\begin{eqnarray}
1 \ge \sum_{n\in I'}\abs{\mel{n}{S}{a}}^2=\sum_{n\in I'}\mel*{a}{S^\dagger}{n}\mel{n}{S}{a} 
\label{eq:unitarity01}
\end{eqnarray}

We can further remove the unit operator from the S-matrix and obtain the scattering amplitude $iM$
\begin{eqnarray}
S = \mathbf{1} + iM
\end{eqnarray}
Provided there are some eigenstate $\ket{\sigma}$ of $M$ with eigenvalue $\sigma$ (i.e. $M\ket{\sigma}=\sigma\ket{\sigma}$), 
then in the subspace we consider, the condition (\ref{eq:unitarity01}) will be transferred to
\begin{eqnarray}
\mel*{\sigma}{S^\dagger S}{\sigma} = \abs{1+i\sigma}^2 \le 1  
\end{eqnarray}
Then the unitary bound on the eigenvalue $\sigma$ is simply
\begin{eqnarray}
    \abs{\sigma}\le 2\,,\quad\quad \abs{\Re(\sigma)}\le 1  \, .
\end{eqnarray}
What we need to do is to find the eigenvalues of $M$ in a suitable subspace. 

Now consider $2\to 2$ scattering process with initial state $\ket{\mathbf{p}_1,\mathbf{p}_2;\lambda_1,\lambda_2}$ and final state $\ket{\mathbf{p}'_1,\mathbf{p}'_2;\lambda'_1,\lambda'_2}$, with $\mathbf{p}_i$ and $\lambda_i$ ($\mathbf{p}'_i$ and $\lambda'_i$) the 3-momentums and helicities of initial state particles (final state particles) respectively. 
We need to transform the initial and final states so that we can discuss the eigenstates in a subspace more easily
\begin{itemize}
\item Firstly, both the initial state and final state can be decomposed into the total momentum part and the relative momentum part
\begin{eqnarray}
\ket{\mathbf{p}_1,\mathbf{p}_2;\lambda_1,\lambda_2} &=& \ket{p_\mathrm{t}}\otimes\ket{\mathbf{p}_\mathrm{r};\lambda_1,\lambda_2}   \\ 
\ket{\mathbf{p}'_1,\mathbf{p}'_2;\lambda'_1,\lambda'_2} &=& \ket{p'_\mathrm{t}}\otimes\ket{\mathbf{p}'_\mathrm{r};\lambda'_1,\lambda'_2}
\end{eqnarray}
with
\begin{eqnarray}
p_\mathrm{t} = ( E_1 + E_2 , \mathbf{p}_1 + \mathbf{p}_2 ) \ , \  \mathbf{p}_\mathrm{r} = \mathbf{p}_1 = - \mathbf{p}_2  \\ 
p'_\mathrm{t} = ( E'_1 + E'_2 , \mathbf{p}'_1 + \mathbf{p}'_2 ) \ , \  \mathbf{p}'_\mathrm{r} = \mathbf{p}'_1 = - \mathbf{p}'_2
\end{eqnarray}
And it should be emphasized that we are working on the center-of-mass frame.
\item Secondly, for a certain central energy $E_{\rm cm}$, we can use the direction of motion of the first particle $(\theta,\phi)$ to label the 
initial state or final state
\begin{eqnarray}
\ket{\theta,\phi;\lambda_1,\lambda_2} = \frac{1}{2\pi}\sqrt{\frac{p_\mathrm{r}}{4E_{\rm cm}}}\ket{\mathbf{p}_\mathrm{r};\lambda_1,\lambda_2}  
\end{eqnarray}
with $p_\mathrm{r} \equiv \abs{\mathbf{p}_\mathrm{r}} $.
For a detailed explanation of the coefficient on the right side of above equation, see Appendix \ref{appA}.
Using the representation theory of angular momentum operator, we can decompose state $\ket{\theta,\phi;\lambda_1,\lambda_2}$ by the eigenstates of angular momentum 
\begin{eqnarray}
\ket{\theta,\phi,\lambda_1,\lambda_2}= \sum_{j=\abs{\lambda}}^\infty\sum_{m=-j}^j \sqrt{\frac{2j+1}{4\pi}} D^j_{m\lambda}(\phi,\theta,0)\ket{j,m;\lambda_1,\lambda_2}
\end{eqnarray}
where $D^j_{m\lambda}$ is the matrix element of the $(2j+1)$-dimensional irreducible representation of the rotation group, and $\lambda \equiv \lambda_1 - \lambda_2$ is the total helicity of initial state. 
Final state can also be decomposed like that. 
\end{itemize}

Now we can choose the frame wisely to simplify our calculation. Firstly, we can choose the direction of particle 1 in the initial stats as $z$-axis, and thus the initial state become
\begin{eqnarray}
\ket{i} \equiv 2\pi  \sqrt{\frac{4E_{\rm cm}}{p_\mathrm{r}}} \ket{p_\mathrm{t}}\otimes  \ket{0,0;\lambda_1,\lambda_2} 
\end{eqnarray}
Due to the axisymmetry of scattering process, we can choose the final state particles to have $\phi=0$, and thus the final state become
\begin{eqnarray}
\ket{f} \equiv 2\pi  \sqrt{\frac{4E_{\rm cm}}{p'_\mathrm{r}}} \ket{p'_\mathrm{t}}\otimes  \ket{\theta,0;\lambda'_1,\lambda'_2} 
\end{eqnarray}
with $\theta$ the scattering angle.

As we said before, we need to find the eigenvalue of $M$ in a suitable subspace. 
Due to the conservation of total momentum and the total angular momentum, we can consider the subspace with certain $p_\mathrm{t}$ or $j,m$.
Then the scattering amplitude can be decomposed as
\begin{eqnarray}
\mel{f}{M}{i} = (2\pi)^4 \delta^4( p_\mathrm{t} - p'_\mathrm{t} ) \mel{f}{ M_{p_\mathrm{t}} }{i}
\end{eqnarray}
with $M_{p_\mathrm{t}}$ the scattering operator in the subspace with total momentum $p_\mathrm{t}$. 
$\mel{f}{ M_{p_\mathrm{t}} }{i}$ is actually the amplitude obtained by Feynman rules (also labeled as $\mathcal{M}$). 
We can further decompose the amplitude by the eigenstates of angular momentum
\begin{eqnarray}
\mathcal{M} \equiv \mel{f}{M_{p_\mathrm{t}}}{i} &=&  (2\pi)^2\frac{4E_{\rm cm}}{\sqrt{p_\mathrm{r} p'_\mathrm{r}}}\mel*{\theta,0,\lambda'_1,\lambda'_2}{M_{p_\mathrm{t}}}{0,0,\lambda_1,\lambda_2} \\\nonumber
&=&   (2\pi)^2\frac{4E_{\rm cm}}{\sqrt{p_\mathrm{r} p'_\mathrm{r}}}  \sum_{j=\abs{\lambda}}^\infty\sum_{m=-j}^j \sum_{j'=\abs{\lambda'}}^\infty\sum_{m'=-j'}^{j'} \sqrt{\frac{2j'+1}{4\pi}} \sqrt{\frac{2j+1}{4\pi}}  \\\nonumber
& &\times (D^{j'}_{m'\lambda'}(0,\theta,0))^\ast D^j_{m\lambda}(0,0,0) \braket{j',m'}{j,m} \mel*{\lambda'_1,\lambda'_2}{M^j_{p_\mathrm{t}}}{\lambda_1,\lambda_2} \\\nonumber
&=& \frac{4\pi E_{\rm cm}}{\sqrt{p_\mathrm{r} p'_\mathrm{r}}}\sum_{j=\max(\abs{\lambda},\abs{\lambda'})}^\infty(2j+1)d^j_{\lambda\lambda'}(\theta) \mel*{\lambda'_1,\lambda'_2}{M^j_{p_\mathrm{t}}}{\lambda_1,\lambda_2}
\end{eqnarray}

By using the orthogonality of $d^j_{\lambda\lambda'}(\theta)$, we have
\begin{eqnarray}
\mel*{\lambda'_1,\lambda'_2}{M^j_{p_\mathrm{t}}}{\lambda_1,\lambda_2}=\frac{\sqrt{p_\mathrm{r} p'_\mathrm{r}}}{8\pi E_{\rm cm}}\int_0^\pi  \mathcal{M} \   d^j_{\lambda\lambda'}(\theta)\sin\theta\dd{\theta}\,.
\end{eqnarray}
Therefore, after calculating amplitude $\mathcal{M}$ by using Feynman rules, we can obtain the matrix elements of $M$ in the subspace with certain $p_\mathrm{t}$ and $j$. 
And the unitary bound simply require the real part of $M$ eigenvalue in this subspace to be smaller than 1. 

Finally, if there are $n$ pairs of identical particles in the initial state and final state, above equation should be modified as
\begin{equation}
\mel*{\lambda'_1,\lambda'_2}{M^j_{p_\mathrm{t}}}{\lambda_1,\lambda_2}=\left(\frac{1}{\sqrt{2}}\right)^n \frac{\sqrt{p_\mathrm{r} p'_\mathrm{r}}}{8\pi E_{\rm cm}}\int_0^\pi \mathcal{M} \  d^j_{\lambda\lambda'}(\theta)\sin\theta\dd{\theta}\,. \label{eq:PartialWaveAmp}
\end{equation}
The detailed derivation process is shown in Appendix \ref{appA}.

\subsection{Upper bound on dark Higgs mass \label{subsec:UpperBound} }

In our model, there are three types of physical particles: dark matter particle $\chi$, dark photon $A'$, and dark Higgs $s$. 
Due to the lack of Yukawa coupling in our model, dark Higgs mass $m_s$ or the Higgs field self-coupling constant $\lambda$ will not be involved in tree-level $2\to 2$ process with $\chi$ in initial or final states. 
So we can ignore the processes related to $\chi$, and focus on $A'$ and $s$. 
 For the scalar particle $s$, the helicity is simply 0. 
 For massive vector boson $A'$, there are three helicity eigenstates. 
 If we choose the momentum of $A'$ to be $\mathbf{p}=\abs{\mathbf{p}}\mathbf{e}_z$, then the polarization vectors for helicity eigenstates are given by
\begin{eqnarray}
\epsilon^+=\begin{pmatrix}
0 \\ 1 \\ i \\ 0
\end{pmatrix};\quad \epsilon^-=\begin{pmatrix}
0 \\ 1 \\ -i \\ 0
\end{pmatrix}; \quad \epsilon^L=\begin{pmatrix}
\frac{\abs{\mathbf{p}}}{m_{A'}} \\ 0 \\ 0 \\ \frac{E}{m_{A'}}
\end{pmatrix}\,.
\end{eqnarray}


For an arbitrary momentum $p^\mu=(E,\pm\abs{\mathbf{p}}\sin\theta,0,\pm\abs{\mathbf{p}}\cos\theta)$ on the $xz$-plane, the polarization vectors are given by
\begin{eqnarray}
\epsilon^+(p)=\begin{pmatrix}
0 \\ \cos\theta \\ \pm i \\ -\sin\theta
\end{pmatrix};\quad \epsilon^-(p)=\begin{pmatrix}
0 \\ \cos\theta \\ \mp i \\ -\sin\theta
\end{pmatrix}; \quad \epsilon^L(p)=\begin{pmatrix}
\frac{\abs{\mathbf{p}}}{m_{A'}} \\ \pm\frac{E}{m_{A'}}\sin\theta \\ 0 \\ \pm\frac{E}{m_{A'}}\cos\theta
\end{pmatrix}\,.    
\end{eqnarray}


All the tree-level $2\to 2$ processes involving only $A'$ and $s$ are 
\begin{align}
A'_{\pm,L}A'_{\pm,L} &\to A'_{\pm,L}A'_{\pm,L}\,;\\
ss &\to ss\,;\\
A'_{\pm,L}A'_{\pm,L} &\to ss\,;\\
ss &\to A'_{\pm,L}A'_{\pm,L}\,;\\
A'_{\pm,L} s &\to A'_{\pm,L} s\,.
\end{align}
Therefore, the scattering amplitudes is divided into two groups.
The initial and final states of one group can only be $ A'_{\pm,L} s$, while the initial and final states of another group can only be $ A'_{\pm,L} A'_{\pm,L}$ or $ss$.

The scattering amplitudes for $ A'_{\pm,L} s \to A'_{\pm,L} s$ under unitary gauge ($\xi=\infty$) is
\begin{align}
i\mathcal{M} &= i 2 g^{\prime 2}\epsilon^\mu(p_1)\epsilon_\mu^*(p_3)+i 4g^{\prime 4}v_{\rm cl}^2\frac{\epsilon_\mu(p_1)\epsilon_\nu^*(p_3)}{(p_1+p_2)^2-m_{ A'}^2}\left(\eta^{\mu\nu}-\frac{(p_1+p_2)^\mu(p_1+p_2)^\nu}{m_{ A'}^2}\right)\nonumber \\
&\qquad +\,i 4g^{\prime 4}v_{\rm cl}^2\frac{\epsilon_\mu(p_1)\epsilon_\nu^*(p_3)}{(p_1-p_4)^2-m_{ A'}^2}\left(\eta^{\mu\nu}-\frac{(p_1-p_4)^\mu(p_1-p_4)^\nu}{m_{ A'}^2}\right)\nonumber \\
&\qquad +\, i 6 g^{\prime 2}\lambda v_{\rm cl}^2\frac{\epsilon^\mu(p_1)\epsilon_\mu^*(p_3)}{(p_1-p_3)^2-m_s^2}\,.
\end{align}

In order to highlight the terms that increase as the energy scale increases (there may not be such terms) or the constant terms in it, we take the limit of $E_{\rm cm}\to +\infty$, and then we find that the terms proportional to the first and second powers of $E_{\rm cm}$ in the above formula are all zero (this is actually guaranteed by the Higgs mechanism).  
Then the leading terms in the $E_{\rm cm}\to +\infty$ limit are those constant terms
\begin{eqnarray}
\lim_{E_{\rm cm}\to\infty}\mathcal{M}( A' s\to A' s)=
\left( \begin{array}{c|ccc}
& A'_Ls & A'_+s & A'_-s \\ 
\midrule
 A'_Ls & -\lambda-\frac{g^{\prime 2}}{4}\frac{7+\cos(2\theta)}{\cos^2(\theta/2)} & 0 & 0\\
 A'_+s & 0 & -4g^{\prime 2} & 0\\
 A'_-s & 0 & 0 & -4g^{\prime 2}
 \end{array} \right) .  
\end{eqnarray}
Here we use the first row and column to denote the initial and final state.  
In the Higgs mass upper bound estimation, gauge coupling constant $g'$ is generally much smaller than $\lambda$ and thus can be ignored.
Then the only process should be considered in unitarity bound estimation is
\begin{eqnarray}
\lim_{E_{\rm cm}\to\infty}\mathcal{M}( A'_Ls\to A'_Ls) \approx -\lambda = -2\frac{m_s^2}{v_{\rm cl}^2}\,.
\end{eqnarray}

Through similar calculations, we can get the matrix corresponding to the process between $ A' A'$ and $ss$ in the $E_{\rm cm}\to +\infty$ limit as
\small 
\begin{eqnarray}
&& \lim_{E_{\rm cm}\to\infty}\mathcal{M}( A' A',ss \to A' A',ss) = \nonumber \\
&& \quad  \quad  \quad  \quad  \left( \begin{array}{c|ccccccc}
 & A'_L A'_L & A'_L A'_+ & A'_L A'_- & A'_+ A'_+ & A'_+ A'_- & A'_- A'_- & ss\\
 \midrule
 A'_L A'_L & -3\lambda\\
 A'_L A'_+ & 0 & -4g^{\prime 2} \\
 A'_L A'_- &  0 & 0 &  -4g^{\prime 2} \\
 A'_+ A'_+ & 0 & 0 & 0 &  0 \\
 A'_+ A'_- & -4g^{\prime 2} & 0 & 0 & 0 & 0 \\
 A'_- A'_- & 0 & 0 & 0 & 0 & 0 & 0 \\
ss & \lambda-g^{\prime 2}\frac{7+\cos(2\theta)}{\sin^2\theta} & 0 & 0 & 0 & 4g^{\prime 2} & 0 & -3\lambda
 \end{array} \right) .  
\end{eqnarray}
\normalsize 
The elements not written in it can be given by symmetry. If terms proportional to $g^{\prime 2}$ are ignored, then we only need to consider longitudinal polarization and scalar as initial or final state
\begin{eqnarray}
\lim_{E_{\rm cm}\to\infty}\mathcal{M}( A'_L A'_L,ss \to A'_L A'_L,ss) \approx \begin{pmatrix}
-3\lambda & \lambda\\
\lambda & -3\lambda
\end{pmatrix}=2\frac{m_s^2}{v_{\rm cl}^2}\begin{pmatrix}
-3 & 1\\
1 & -3
\end{pmatrix}\,.
\end{eqnarray}

 Next, we apply Eq.~\eqref{eq:PartialWaveAmp} in the $E_{\rm cm}\to \infty$ limit
\begin{eqnarray}
\mel*{\lambda'_1,\lambda'_2}{M^j_{p_\mathrm{t}}}{\lambda_1,\lambda_2}\approx \left(\frac{1}{\sqrt{2}}\right)^n\frac{1}{16\pi}\int_0^\pi \mathcal{M} \ d^j_{\lambda\lambda'}(\theta)\sin\theta\dd{\theta}\,.
\end{eqnarray}
According to current calculation results, in the $E_{\rm cm}\to \infty$ limit we only need to consider states with 0 helicity.
Because $d^l_{00}(\theta)=P_l(\cos\theta)$ and the scattering amplitudes in the $E_{\rm cm}\to \infty$ limit has nothing to do with angular $\theta$, thus all the partial wave matrix elements $j>0$ are equal to 0. 
For $j=0$, we obtain 
\begin{eqnarray}
\lim_{E_{\rm cm}\to\infty}\mel*{0,0}{M^{j=0}}{0,0}=\frac1{8\pi}\frac{m_s^2}{v_{\rm cl}^2}\left( \begin{array}{c|ccc}
  & A'_L A'_L & ss & A'_Ls \\
 \midrule
A'_L A'_L & -3 & 1 & 0\\
ss & 1 & -3 & 0\\
A'_Ls & 0 & 0 & -2
 \end{array} \right) .  
\end{eqnarray}
The above formula has taken into account the $1/\sqrt{2}$ factor caused by identical particles. Among the three eigenvalues of matrix
\begin{eqnarray}
\begin{pmatrix}
-3 & 1 & 0\\
1 & -3 & 0\\
0 & 0 & -2
\end{pmatrix}\,,
\end{eqnarray}
the eigenvalue with the largest absolute value is $-4$, and the corresponding eigenvector is $(1,-1,0)^{\rm T}$. Therefore, for the two-particle state
\begin{eqnarray}
\ket{\psi}=\frac{1}{\sqrt{2}}\ket{j=j_z=0; A'_L A'_L}-\frac{1}{\sqrt{2}}\ket{j=j_z=0;ss}\,,
\end{eqnarray}
we have
\begin{eqnarray}
\lim_{E_{\rm cm}\to\infty}\mel*{\psi}{M}{\psi}=-\frac1{2\pi}\frac{m_s^2}{v_{\rm cl}^2}\,.
\end{eqnarray}
On the other hand, unitarity requires $\abs{\Re(\mel*{\psi}{M}{\psi})}\le 1$, from which we get
\begin{eqnarray}
\label{massuplimit}
m_s^2\le 2\pi v_{\rm cl}^2 \,.
\end{eqnarray}
This is the upper bound of the dark Higgs mass that we finally obtained. 

\section{Lower bound on dark Higgs mass from one-loop effective potential}
\label{sec:lower}

In the SM, the lower bound of Higgs mass comes from the vacuum stability requirement. 
More specifically, the large top quark Yukawa coupling contributes a negative term to the $\beta$ function of Higgs self-coupling.
So if the Higgs mass (which is proportional to self-coupling) is too small, Higgs self-coupling will be less than zero before Planck scale and destabilize the electroweak vacuum~\cite{Degrassi:2012ry,Buttazzo:2013uya,Cabibbo:1979ay,Sher:1988mj,Casas:1994qy,Andreassen:2014gha}. 
The lower bound of the SM Higgs mass from this vacuum stability requirement is actually very close to current measured value~\cite{ATLAS:2023xpr,CMS:2020xrn}. 

\begin{figure}[ht]
\centering
\includegraphics[width=0.5\textwidth]{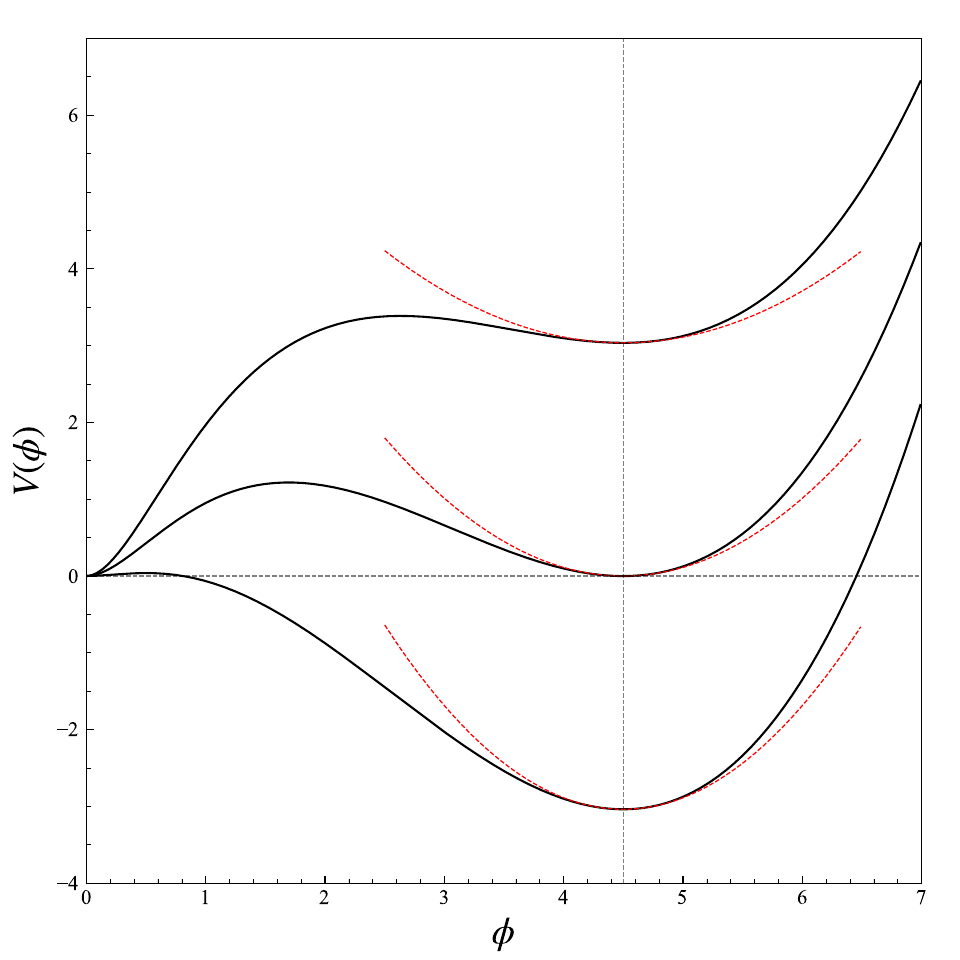}
\caption{ Illustration of the Linde-Weinberg bound. Curvatures of red dotted lines correspond to Higgs mass. }
\label{fig:Illustrate4WeinbergBound}
\end{figure}

In our dark sector model, the dark matter is vector-like fermion charged under $U(1)'$ and thus there is no Yukawa coupling in the dark sector. 
So, the value of $m_s$ can not be bounded from below by dark vacuum stability. 
However, even in this case, there is still a theoretical bound that can be used to limit the lower bound of $m_s$. 
This bound is the Linde-Weinberg bound~\cite{Linde:1975sw,Weinberg:1976pe} which says that $m_s$ can not be too light, otherwise the shape of scalar potential will become a ``bowl'' shape instead of a ``Mexican-hat'' shape and thus the spontaneous symmetry breaking (SSB) cannot actually occur. 
For a rough illustration of the Linde-Weinberg bound, see Fig.~\ref{fig:Illustrate4WeinbergBound}. 
In this figure, we keep the VEV $v$ that causes symmetry breaking constant, and then adjust parameters so that the effective potential $V(v)$ continuously increases. 
Since the mass of the Higgs boson is approximately equal to $V''(v)$, the mass of the Higgs boson decreases as $V(v)$ increases. When $V(v)$ is at the critical position for symmetry breaking, the corresponding Higgs boson mass is the Linde-Weinberg bound.

It should be emphasized that the Linde-Weinberg bound can only appear at loop level. 
If one only consider tree-level potential then there is no lower bound for $m_s$.
In addition, there are other subtleties need to be clarified, which we will discuss in the following subsections.

\subsection{Gauge dependence of effective potential\label{subsec:gauge}}

\begin{figure}[ht]
\centering
\includegraphics[width=4.5in]{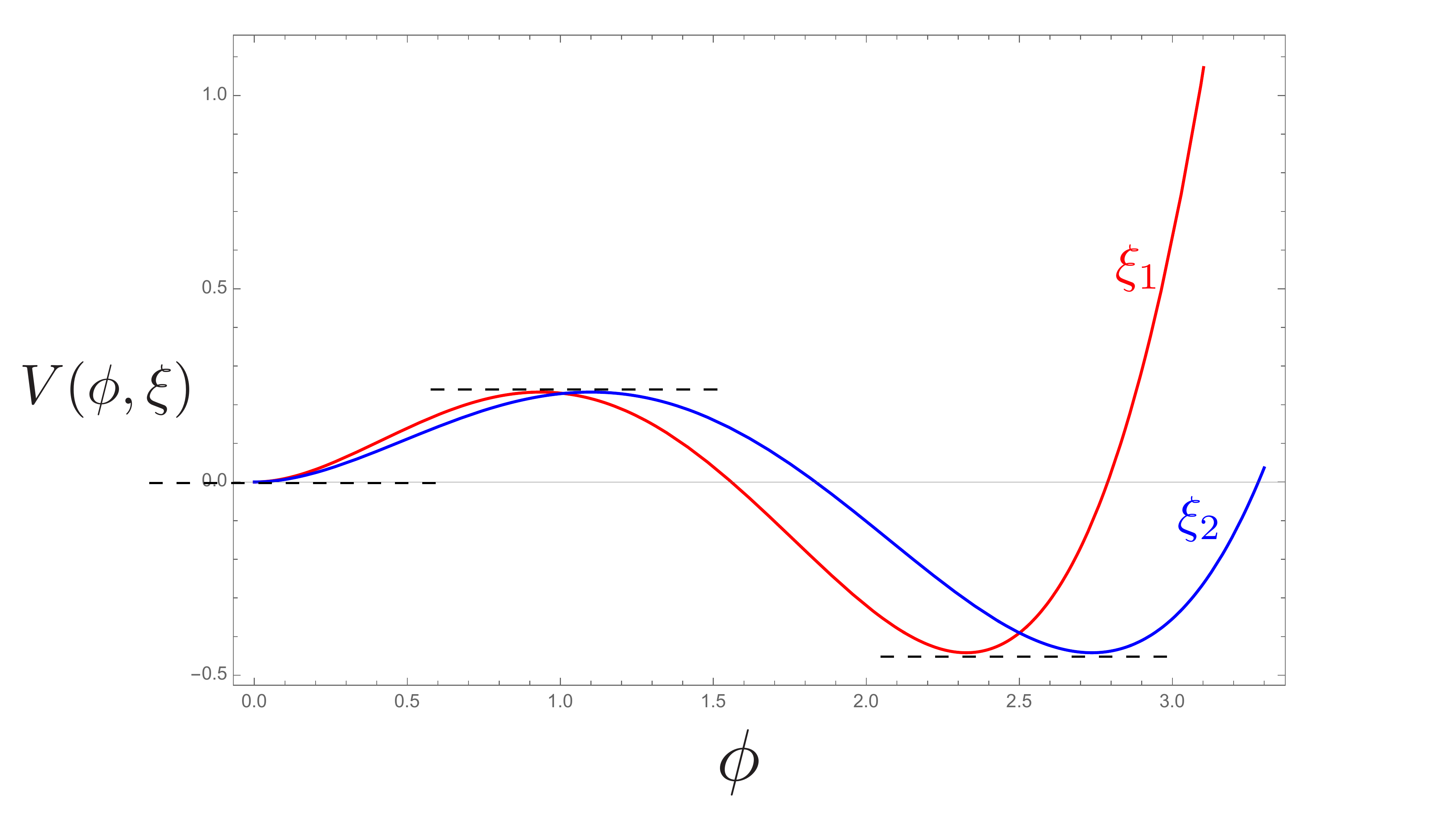}
\caption{ Illustration plot for the gauge dependence of $V(\phi,\xi)$. This plot is similar to Fig.1 in Ref.~\cite{Andreassen:2014eha}.}
\label{fig:IllustrationEffectP}
\end{figure}

As the summation of all 1PI vertices, the effective potential for gauge theory is born to be gauge dependent, i.e. $V(\phi,\xi)$ with $\xi$ being the gauge parameter.
However, all the physical quantities must be gauge independent. So when using $V(\phi,\xi)$ to induce physical quantity like Linde-Weinberg bound, we need to ensure that all the gauge dependencies cancel out eventually.
For illustration, we present a plot in Fig.~\ref{fig:IllustrationEffectP}, which is similar to Fig.1 in Ref.~\cite{Andreassen:2014eha}.
From Fig.~\ref{fig:IllustrationEffectP} we see that the shapes of $V(\phi,\xi_1)$ and $V(\phi,\xi_2)$ are different, but the extreme values are the same. 
This feature is actually required by Nielsen identity~\cite{NIELSEN1975173,PhysRevD.13.3469}:  
\begin{eqnarray}
\left(\xi\frac{\partial}{\partial\xi} + C(\phi,\xi)\frac{\partial}{\partial\phi}   \right)V(\phi,\xi) = 0\,,
\end{eqnarray}
with $C(\phi,\xi)$ being a calculable function. 
We denote the position of $\phi$ which minimize $V(\phi,\xi)$ away from the zero point as $v(\xi)$ ($v$ is certainly gauge dependent). 
For SSB to happen, the following potential difference needs to be positive: 
\begin{eqnarray}
\Delta V\equiv  V(0,\xi) - V(v(\xi),\xi) > 0\,.
\end{eqnarray}
If $\Delta V$ is negative, then SSB can not happen. 
The Linde-Weinberg bound of $m_{s}$ comes from this condition. 

Due to Nielsen identity, $\Delta V$ is clearly gauge independent. 
However, it should be known that Nielsen identity is non-perturbative (for recent elegant proofs of Nielsen identity, see, e.g., ~\cite{DelCima:1999dr,DelCima:1999gg}).   
So, in a specific perturbation calculation, the gauge independence of $\Delta V$ needs to be confirmed order-by-order. 
At one-loop level, this problem has been solved in the literature~\cite{Dolan:1974gu,Loinaz:1997td,Aitchison:1983ns,DoNascimento:1987mn,Andreassen:2014eha,Patel:2011th}.
For the self-consistency of this article, here we make a brief review. 

Assuming we have obtained an one-loop effective potential as
\begin{equation}
V_{\rm 1L}(\phi)=V_0(\phi)+V_1(\phi)\,,
\end{equation}
where $V_1(\phi)$ is proportional to $\hbar$. In the calculations in the following sections, we will explicitly include $\hbar$ to highlight the corresponding number of loops in the Feynman diagrams. When performing numerical calculations, we will set $\hbar=1$. If we directly use the condition satisfied by $v$:
\begin{equation}
\left.\dv{V_{\rm 1L}(\phi)}{\phi}\right|_{\phi=v}=0\,,
\end{equation}
then the obtained $v$ will include partial two-loop corrections, and thus $V_{\rm 1L}(v)$ will also contain partial two-loop corrections, making it gauge-dependent in general. Only when ignoring these incomplete two-loop corrections can the obtained $V_{\rm 1L}(v)$ become gauge-independent. For this end, we set
\begin{equation}
v=v_{\rm cl}+v_1+\mathcal{O}(\hbar^2)\,,
\end{equation}
and then strictly expanding in $\hbar$ gives
\begin{equation}\label{eq:Vmin1LExpand}
\begin{aligned}
V(v) &= V_0(v_{\rm cl})+v_1V'_0(v_{\rm cl})+V_1(v_{\rm cl})+\mathcal{O}(\hbar^2)\\
&= V_0(v_{\rm cl})+V_1(v_{\rm cl})+\mathcal{O}(\hbar^2)\,,
\end{aligned}
\end{equation}
where we used the relation $V_0'(v_{\rm cl})=0$ satisfied by $v_{\rm cl}$. The Nielsen identity ensures that $V_0(v_{\rm cl})+V_1(v_{\rm cl})$ is gauge-independent.
We will also prove the gauge-independence in Sec.~\ref{subsec:oneloopVeff} via a concrete calculation.

\subsection{Renormalization-scale dependence of effective potential \label{subsec:renormalize}}

Another issue we face is the dependence of the effective potential on the renormalization scale $\mu_r$. 
As fictitious parameter when performing $\overline{\text{MS}}$ renormalization scheme, the value of $\mu_r$ should not affect effective potential.  
To make this point clear, consider the generating functional of connected diagrams $W[J]$: 
\begin{equation}
W[J] = -i \log Z[J] = -i \log \int\mathcal{D}\phi\,e^{i\int\dd{}^dx\left(\mathcal{L}\,+\,\phi J\right)} \, .
\end{equation}
Apparently $W[J]$ should be independent of the fictitious scale $\mu_r$:
\begin{equation}
\mu_r\dv{\mu_r}W[J] = 0 \, .
\end{equation}
As the Legendre transformation of $W[J]$, the effective potential should also be independent of $\mu_r$:
\begin{equation}
\mu_r\dv{\mu_r} V(\phi,g_i,\mu_r) = 0 \, ,
\end{equation}
where $g_i$ (can be either masses or couplings) are renormalized parameters in the Lagrangian. And since we use $\overline{\text{MS}}$ scheme, $V$ is an explicit function of $\mu_r$. 

In some literature, when using effective potential $V$ to study the SSB of the theory, they simply regard renormalized parameters as physical parameters (for example, regard the renormalized mass in the Lagrangian as the physical mass). 
But this method will lead to non-physical dependence of $V$ on $\mu_r$. 
This is because the physical parameters are $\mu_r$ independent by themselves. 
In Sec.~\ref{subsec:ConstraintsBySSB} we will show that such a non-physical scale dependence will lead to unreliable results.

To overcome this problem, a common approach in the literature is to use the RGE (renormalization group equation) running of parameters to offset the explicit dependence on $\mu_r$:
\begin{equation}
\mu_r\dv{\mu_r} \left. V(\phi,g_i,\mu_r)\right|_{\phi=v} = \mu_r\pdv{\mu_r} \left. V(\phi,g_i,\mu_r)\right|_{\phi=v} + \sum_i\beta_i\pdv{g_i} \left. V(\phi,g_i,\mu_r)\right|_{\phi=v} = 0 
\end{equation}
with $\beta_i \equiv d g_i / d \ln \mu_r$ being the beta function (or anomalous dimension), and $\phi=v$ being the position that minimize $V$. 
In practical calculations, the dependence on $\mu_r$ cannot be fully eliminated due to the truncation of higher-order corrections.

In this work, we solve the $\mu_r$-dependence problem by a different method. 
We will try to re-express the effective potential by using physical parameters (or say physical observables) rather than renormalized parameters. 
To be more specific, at one-loop level, the renormalized parameters $g_i$ can be expressed by physical parameters $g_{i,p}$: 
\begin{equation}
\label{onelooprelation}
g_i = g_{i,p} + \hbar G_i (g_{j,p},\mu_r)
\end{equation}
where $G_i$ are the functions from one-loop diagrams. 
Then the minimum of one-loop effective potential, which is originally expressed by $g_i$, can be re-expressed by $g_{i,p}$, with difference at $\mathcal{O}(\hbar^2)$ order:
\begin{equation}
\left. V_{\text{1L}}(g_i,\mu_r)\right|_{\text{min}} = \left. V_{p,\text{1L}}(g_{p,i})\right|_{\text{min}} + \mathcal{O}(\hbar^2) \, .
\end{equation}
Here we use $V_{p}$ to label the effective potential expressed by physical parameters.
Note that the minimum of effective potential is not a function of field value $\phi$ anymore. 
The key point here is that $\left. V_{p,\text{1L}}(g_{p,i})\right|_{\text{min}}$ no longer explicitly contains the scale $\mu_r$. 
This is simply because $\left. V_{p,\text{1L}}\right|_{\text{min}}$ is expressed by physical parameters which are independent of the scale $\mu_r$. 
In Sec.~\ref{subsec:Re-parameterize} we will calculate the one-loop relationship Eq.~(\ref{onelooprelation}) in our model, and clearly prove that $\left. V_{p,\text{1L}}\right|_{\text{min}}$ has nothing to do with the scale $\mu_r$.

In our current dark $U(1)'$ model, the dark matter field $\chi$ does not directly couple with the dark Higgs, so the parameters related to $\chi$ do not appear in the one-loop effective potential. 
Thus, the one-loop effective potential depends on three parameters, which are conventionally taken as $\{\mu,\lambda,g'\}$ or $\{g',m^2_s \equiv 2\mu^2, m^2_{A'}  \equiv g'^2 (4\mu^2/\lambda)\}$. 
As we explained above, expressing the effective potential by renormalized parameters such as $\{\mu,\lambda,g'\}$ makes the SSB criterion ($\Delta V > 0$) scale dependent.
To obtain $\Delta V$ that does not explicitly contain $\mu_r$, we will finally re-parameterize the effective potential with the following three physical parameters:
\begin{equation}
m_{A',p}\,,\quad m_{s,p}\,,\quad g'_p\,,
\end{equation}
where $m_{A',p}$ and $m_{s,p}$ are the pole masses of the dark photon and dark Higgs, respectively, and $g'_p$ is defined by the scattering amplitude of the DM particle $2\to2$ process at transfer momentum $q^2=-m_{A',p}^2$. 
We choose these three physical parameters as input for convenience in Sec.~\ref{sec:fit}.  

In the following two subsections, we will solve the problem of gauge dependence and $\mu_r$-dependence of the SSB criterion in our model.

\subsection{The one-loop-level effective potential and \texorpdfstring{$\Delta V$}{Delta V}\label{subsec:oneloopVeff}}

To calculate the effective potential, we write the field $S$ as
\begin{equation}
S=\frac{1}{\sqrt{2}}(s+\eta_1+i(a+\eta_2))\,,
\end{equation}
where $s$ and $a$ are the vacuum expectation values of the dark Higgs field and the Goldstone field, respectively.
To fix the gauge, one can start from the most general 3-parameter gauge:
\begin{equation}
\mathcal{L}_{\mathrm{gf}} = -\frac{1}{2 \xi}\left(\partial_\mu A'^\mu+ g' \xi \Upsilon_1 (s+\eta_1)+ g' \xi \Upsilon_2 (a+\eta_2)\right)^2 \, .
\end{equation}
We can use the $U(1)'$ symmetry to keep $a=0$, then choose $\Upsilon_1 = v_{\rm cl}$ and $\Upsilon_2  = 0$ as suggested in the literature~\cite{Dolan:1974gu,Fukuda:1975di,DoNascimento:1987mn}. 


Since we aim to compute the effective potential at the one-loop level, we only need to consider the quadratic terms of the fields in the Lagrangian:
\begin{equation}\label{eq:quadraticINL}
\begin{aligned}
\mathcal{L}_{\text{quadratic}} &= -\frac{1}{4}F_{\mu\nu}'^2+\frac12m_{A'}^2(s)A'_\mu A'^\mu-\frac1{2\xi}(\partial_\mu A'^\mu)^2\\
&\qquad +\,\frac12\left((\partial_\mu\eta_1)^2+(\partial_\mu\eta_2)^2\right)-\frac12m_s^2(s)\eta_1^2 - \frac12m_a^2(s)\eta_2^2  \\
&\qquad +\,\partial_\mu\bar{c}\,\partial^\mu c-m_G^2(s)\bar{c}c+g'(s-v_{\rm cl})A'^\mu\partial_\mu\eta_2\,,
\end{aligned}
\end{equation}
where the field-dependent masses are defined as
\begin{eqnarray}
&& m_{A'}^2(s) = g'^2s^2\,,\\
&& m_s^2(s) = -\mu^2+\frac34\lambda s^2\,,\\
&& m_a^2(s) = -\mu^2+\frac{\lambda}{4}s^2+\xi g'^2v_{\rm cl}^2\,,\\
&& m_G^2(s) = \xi g'^2v_{\rm cl}s\,.
\end{eqnarray}

With the help of Eq.~\eqref{eq:quadraticINL}, we can obtain
\small
\begin{eqnarray}\label{eq:DeterminantExpress4V1}
e^{-iL^4 V_{1}(s)}
&=&\,\int\prod_\alpha\mathcal{D}\phi_\alpha\, e^{-\frac12 A'_\mu iD^{\mu\nu}_{A'}A'_\nu+A'^\mu iK_\mu\eta_2-\frac12\eta_1iD_{\eta_1}\eta_1-\frac12\eta_2iD_{\eta_2}\eta_2-\bar{c}(iD_G)c}\nonumber \\
&=&\,\mathcal{N}_1\mathrm{Det}(iD_G)\left(\mathrm{Det}\left(iD_{A'}^{\mu\nu}\right)\right)^{-1/2}\left(\mathrm{Det}\left(iD_{\eta_1}\right)\right)^{-1/2}\int\mathcal{D}\eta_2\,e^{-\frac12\eta_2\left(iD_{\eta_2}-K_\mu(iD_{A'}^{-1})^{\mu\nu}K_\nu\right)\eta_2}\nonumber  \\
&=&\,\mathcal{N}\mathrm{Det}(iD_G)\left(\mathrm{Det}\left(iD_{A'}^{\mu\nu}\right)\right)^{-1/2}\left(\mathrm{Det}\left(iD_{\eta_1}\right)\right)^{-1/2}\left(\mathrm{Det}\left(iD_{\eta_2}-K_\mu(iD_{A'}^{-1})^{\mu\nu}K_\nu\right)\right)^{-1/2}\,,
\end{eqnarray}
\normalsize
where $V_1(s)$ is the one-loop effective potential and $L^4$ is the volume of whole spacetime.

In momentum space, the operator determinants in Eq.~\eqref{eq:DeterminantExpress4V1} can be straightforwardly calculated. It is important to note that the leftmost $K_\mu$ in $K_\mu(iD_{A'}^{-1})^{\mu\nu}K_\nu$ acts to the left, thus, when transformed into momentum space, it becomes $K_\mu(-p)(iD_{A'}^{-1})^{\mu\nu}K_\nu(p)$. After some algebraic manipulations and using the $\overline{\rm MS}$ scheme, we can obtain
\small
\begin{eqnarray}
V_1(s) &=& \frac{\hbar}{16\pi^2}\left[\frac{3}{4}m_{A'}^4(s)\left(\ln\frac{m_{A'}^2(s)}{\mu_r^2}-\frac{5}{6}\right)+\frac{1}{4}m_s^4(s)\left(\ln\frac{m_s^2(s)}{\mu_r^2}-\frac{3}{2}\right)
\right. \nonumber \\
&&-\frac{1}{2}m_G^4(s)\left(\ln\frac{m_G^2(s)}{\mu_r^2}-\frac{3}{2}\right)+\left.\frac{1}{4}m_+^4(s)\left(\ln\frac{m_+^2(s)}{\mu_r^2}-\frac{3}{2}\right)+\frac{1}{4}m_-^4(s)\left(\ln\frac{m_-^2(s)}{\mu_r^2}-\frac{3}{2}\right)\right]\,,
\end{eqnarray}
\normalsize
where
\begin{equation}
m_\pm^2 = \frac{1}{2}\left(\frac{\lambda}{4}s^2-\mu^2\right)+\xi g'^2v_{\rm cl}s\pm\frac12\sqrt{\left(\frac{\lambda}{4}s^2-\mu^2\right)\left(\frac{\lambda}{4}s^2-\mu^2+4\xi g'^2(v_{\rm cl}s-s^2)\right)}\,.
\end{equation}
This result is consistent with Ref.~\cite{Andreassen:2014eha}. Note that $m_+^2(v_{\rm cl})=m_-^2(v_{\rm cl})=m_G^2(v_{\rm cl})$, therefore we have 
\begin{equation}
V_1(v_{\rm cl}) = \frac{\hbar}{16\pi^2}\left[\frac{3}{4}m_{A'}^4\left(\ln\frac{m_{A'}^2}{\mu_r^2}-\frac{5}{6}\right)+\frac{1}{4}m_s^4\left(\ln\frac{m_s^2}{\mu_r^2}-\frac{3}{2}\right)\right]\,.
\end{equation}
On the other hand, we have
\begin{equation}
V_0(v_{\rm cl})=-\frac{\lambda}{16}v_{\rm cl}^4=-\frac{m_{A'}^2m_s^2}{8g'^2}\,.
\end{equation}
According to Eq.~\eqref{eq:Vmin1LExpand}, we can obtain
\begin{equation}
V(v)=-\frac{m_{A'}^2m_s^2}{8g'^2}+\frac{\hbar}{16\pi^2}\left[\frac{3}{4}m_{A'}^4\left(\ln\frac{m_{A'}^2}{\mu_r^2}-\frac{5}{6}\right)+\frac{1}{4}m_s^4\left(\ln\frac{m_s^2}{\mu_r^2}-\frac{3}{2}\right)\right]+\mathcal{O}(\hbar^2)\,.
\end{equation}
It is evident that $V(v)$ is indeed gauge-independent at one-loop level. Interestingly, when $s=v_{\rm cl}$, only the physical degrees of freedom with non-zero mass contribute to $V_1(v_{\rm cl})$. This is also true for the extremum at $s=0$, where there are only two physical degrees of freedom with non-zero mass, both being scalar degrees of freedom, allowing us to immediately write
\begin{equation}
\begin{aligned}
V(0) &= \frac{\hbar}{16\pi^2}\left[\frac{1}{4}m_s^4(0)\left(\ln\frac{m_s^2(0)}{\mu_r^2}-\frac32\right)+\frac{1}{4}m_s^4(0)\left(\ln\frac{m_s^2(0)}{\mu_r^2}-\frac32\right)\right]\\
&= \frac{\hbar}{16\pi^2}\times\frac{1}{8}m_s^4\left(\ln\frac{-m_s^2}{2\mu_r^2}-\frac32\right)\,.
\end{aligned}
\end{equation}
It is important to note that $V(0)$ is a complex number, and the potential energy in this position correspond to its real part (See Ref.~\cite{Fujimoto:1982tc,Weinberg:1987vp} for more discussion).
Thus we obtain
\begin{equation}
\begin{aligned}
\Delta V &= \Re(V(0))-V(v)\\
&= \frac{m_{A'}^2m_s^2}{8g'^2}-\frac{\hbar}{16\pi^2}\left[\frac{3}{4}m_{A'}^4\left(\ln\frac{m_{A'}^2}{\mu_r^2}-\frac{5}{6}\right)+\frac{1}{8}m_s^4\left(\ln\frac{2m_s^2}{\mu_r^2}-\frac{3}{2}\right)\right]+\mathcal{O}(\hbar^2)\,.
\end{aligned}\label{eq:DeltaVTreeLevelPara}
\end{equation}
Therefore, we obtain the one-loop level condition for symmetry breaking as
\begin{equation}
\frac{m_{A'}^2m_s^2}{8g'^2}-\frac{\hbar}{16\pi^2}\left[\frac{3}{4}m_{A'}^4\left(\ln\frac{m_{A'}^2}{\mu_r^2}-\frac{5}{6}\right)+\frac{1}{8}m_s^4\left(\ln\frac{2m_s^2}{\mu_r^2}-\frac{3}{2}\right)\right]>0\,.
\end{equation}
This result, although gauge-independent, is dependent on the renormalization scale. 
In the next subsection, we will replace the renormalized parameters by physical parameters to remove the scale dependence.

\subsection{Re-parameterize the effective potential
with physical parameters\label{subsec:Re-parameterize}}

The purpose of this subsection is to replace the renormalized parameters $\{ m_{A'}\,, m_{s}\,, g' \}$ with the physical parameters $\{ m_{A',p}\,, m_{s,p}\,, g'_p \}$, at one-loop level.
To do this, we need to calculate dark photon pole mass, dark Higgs pole mass, and scattering cross section between DMs, at one-loop level.
For the conciseness of main text, we put all the calculation details into Appendix \ref{appB}. 

Here we directly present the one-loop level relations. One-loop level relation for dark photon mass is
\small
\begin{equation}
\begin{aligned}
m_{A'}^2 &= m_{A',p}^2-\hbar\Re(\Pi_{A'A'}^{\perp}(m_{A',p}^2))\\
&= m_{A',p}^2+\frac{\hbar g'^2}{8\pi^2}\Re\Bigg[\frac{31 m_{A',p}^2}{9}-\frac{5 m_{s,p}^2}{2}-\frac{m_{A',p}^4}{m_{s,p}^2}+\frac{m_{s,p}^4}{6 m_{A',p}^2}+m_{s,p} \sqrt{4 m_{A',p}^2-m_{s,p}^2}\,\\
&\qquad \times\left(\frac{2 m_{s,p}^2}{3 m_{A',p}^2}-\frac{m_{s,p}^4}{6 m_{A',p}^4}-2\right) \Bigg(\arctan\frac{m_{s,p}}{\sqrt{4 m_{A',p}^2-m_{s,p}^2}}+\arctan\frac{2 m_{A',p}^2-m_{s,p}^2}{\sqrt{4 m_{A',p}^2 m_{s,p}^2-m_{s,p}^4}}\Bigg)\\
&\qquad +\,\left(\frac{m_{s,p}^6}{12 m_{A',p}^4}-\frac{m_{s,p}^4}{2 m_{A',p}^2}\right) \ln\frac{m_{A',p}^2}{m_{s,p}^2}+\left(\frac{3 m_{A',p}^4}{m_{s,p}^2}-\frac{5 m_{A',p}^2}{3}+\frac{3 m_{s,p}^2}{2}\right) \ln\frac{m_{A',p}^2}{\mu ^2}\Bigg]\,.
\end{aligned}\label{eq:mAtomAp}
\end{equation}
\normalsize 
One-loop level relation for dark Higgs mass is
\begin{equation}
\begin{aligned}
m_s^2 &= m_{s,p}^2-\hbar\Re(\Pi_{ss}(m_{s,p}^2))\\
&= m_{s,p}^2-\hbar \frac{g_p'^2}{32\pi^2}\frac{m_{s,p}^2}{m_{A',p}^2}\Re\Bigg[10m_{A',p}^2-12\frac{m_{A',p}^4}{m_{s,p}^2}+\left(3\sqrt{3}\pi-14\right)m_{s,p}^2\\
&\qquad +\, 2m_{s,p}\sqrt{4 m_{A',p}^2-m_{s,p}^2}\left(1-4\frac{m_{A',p}^2}{m_{s,p}^2}+12\frac{m_{A',p}^4}{m_{s,p}^4}\right)\arctan\frac{m_{s,p}}{\sqrt{4 m_{A',p}^2-m_{s,p}^2}}\\
&\qquad\left.+\,(m_{s,p}^2-6m_{A',p}^2)\ln\frac{m_{A',p}^2}{\mu_r^2}+3 m_{s,p}^2\ln\frac{m_{s,p}^2}{\mu_r^2}\right]\,.
\end{aligned}\label{eq:mstomsp}
\end{equation}
One-loop level relation for gauge coupling is
\begin{equation}\label{eq:gtogp}
g'^2=g'^2_p\left(1+\frac{\hbar}{2 m_{A',p}^2}\left(\Re\Pi_{A'A'}^\perp(-m_{A',p}^2)-\Re\Pi_{A'A'}^\perp(m_{A',p}^2)\right)\right)+\mathcal{O}(\hbar^2)\,.
\end{equation}
with $\Pi_{A'A'}^\perp$ and $\Pi_{ss}$ the one-loop self-energy for dark photon and dark Higgs respectively. See Appendix \ref{appB} for detailed expressions. 

Substituting Eqs.~(\ref{eq:mAtomAp}, \ref{eq:mstomsp}, \ref{eq:gtogp}) into Eq.~\eqref{eq:DeltaVTreeLevelPara} and expanding to the first order of $\hbar$, we obtain the formula of $\Delta V_{p,\text{1L}}(g'_p,m_{s,p}^2,m_{A',p}^2)$ expressed in terms of physical parameters:
\begin{align}
\label{eq:SSBcondition}
&\Delta V_{p,\text{1L}}(g'_p,m_{s,p}^2,m_{A',p}^2)\notag\\
=& \,\frac{m_{A',p}^2 m_{s,p}^2}{8 g_p^2}+\frac{\hbar}{768 \pi^2 m_{A',p}^4 m_{s,p}}\Re\Bigg\{\left(25-9 \sqrt{3} \pi -6 \ln2\right)m_{A',p}^4 m_{s,p}^5+4 m_{A',p}^6 m_{s,p}^3 \notag\\
&\qquad +\, 54 m_{A',p}^8 m_{s,p}-\left(12 m_{A',p}^4-4 m_{A',p}^2 m_{s,p}^2+m_{s,p}^4\right)\sqrt{4
   m_{A',p}^2-m_{s,p}^2}\notag\\
&\qquad \times\Bigg[m_{s,p}^4 \arctan\frac{2
   m_{A',p}^2-m_{s,p}^2}{m_{s,p} \sqrt{4 m_{A',p}^2-m_{s,p}^2}}+\left(6 m_{A',p}^4+m_{s,p}^4\right) \arctan\frac{m_{s,p}}{\sqrt{4 m_{A',p}^2-m_{s,p}^2}}\Bigg]\notag\\
&\qquad +\,m_{s,p}^3 \left(m_{s,p}^4-8
   m_{A',p}^4\right) \sqrt{4 m_{A',p}^4+m_{s,p}^4}\,\mathrm{arctanh}\,\frac{\sqrt{4 m_{A',p}^4+m_{s,p}^4}}{2 m_{A',p}^2+m_{s,p}^2}\notag\\
&\qquad +\left(8 m_{A',p}^6 m_{s,p}^3+3
   m_{A',p}^4 m_{s,p}^5-3 m_{A',p}^2 m_{s,p}^7+m_{s,p}^9\right) \ln\frac{m_{A',p}^2}{m_{s,p}^2}\Bigg\}\,.
\end{align}
From this result, we can see that $\Delta V_{p,\text{1L}}(g'_p,m_{s,p}^2,m_{A',p}^2)$ is not only gauge-independent but also independent of the renormalization scale, which is consistent with the discussions in sec.~\ref{subsec:gauge} and~\ref{subsec:renormalize}. 
In next subsection, we will use this gauge- and scale-independent criterion to study the lower bound of dark Higgs. 

\subsection{Constraints on physical parameters and  tree-level parameters from symmetry breaking conditions\label{subsec:ConstraintsBySSB}}

In the preceding section, we obtained the gauge- and scale-independent SSB condition: 
\begin{equation}
\Delta V_{p,\text{1L}}(g'_p,m_{s,p}^2,m_{A',p}^2)>0\,
\end{equation}
with the expression in Eq.(\ref{eq:SSBcondition}). 
It can be used to obtain the gauge- and scale-independent Linde-Weinberg bound.

To highlight the importance of our scale-independent method, in this section we also test the naive scale-dependent SSB condition. 
To be more specific, one can naively consider to induce parameters $\{g',\lambda,\mu\}$ (or equivalently $\{g',m_s^2,m_{A'}^2\}$) by using the tree-level relation: 
\begin{equation}
m^2_s \equiv 2\mu^2 = m_{s,p}^2 \ , \ m^2_{A'}  \equiv g'^2 \frac{4\mu^2}{\lambda}  = m_{A',p}^2 \ , \ g' = g'_p
\end{equation}
And the corresponding scale-dependent SSB condition is written as:
\begin{equation}
\Delta V_{\text{1L}}(g',m_s^2,m_{A'}^2,\mu_r)>0\,
\end{equation}
with the expression given in Eq.(\ref{eq:DeltaVTreeLevelPara}) ($\mathcal{O}(\hbar^2)$ terms neglected).

\begin{figure}[ht]
\centering
\includegraphics[width=0.45\textwidth]{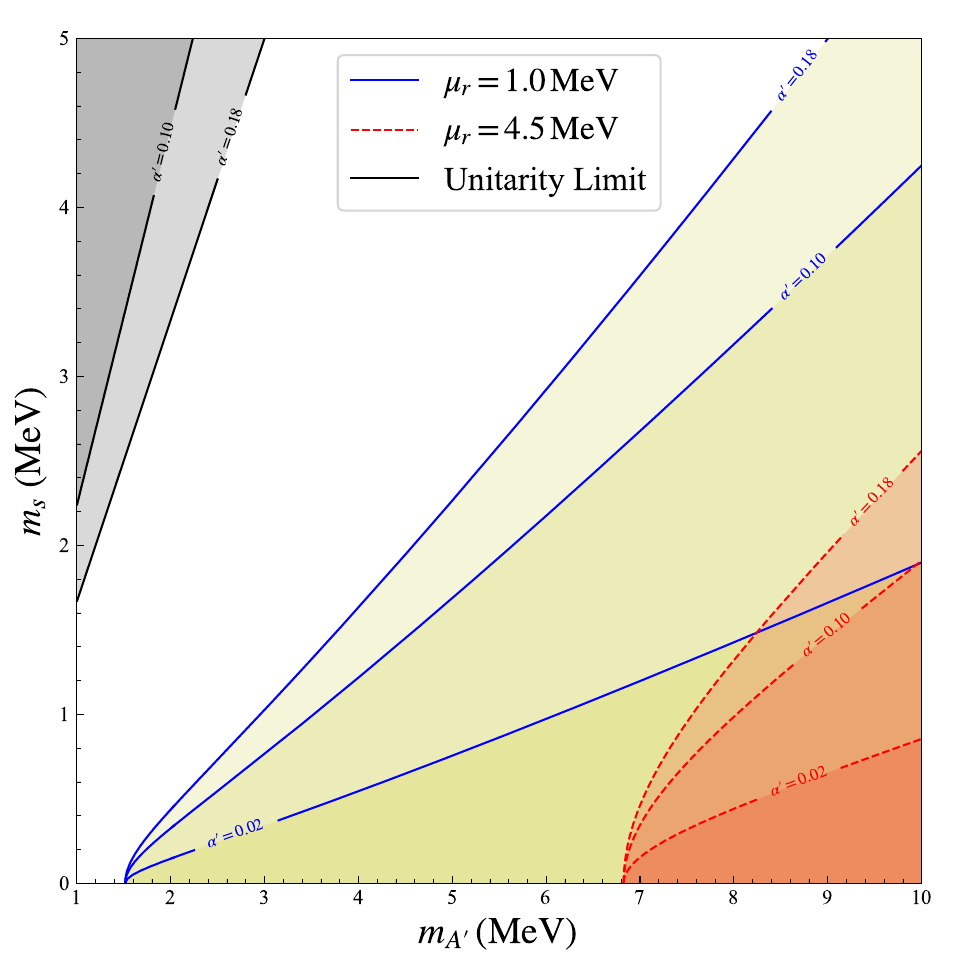}\qquad
\includegraphics[width=0.45\textwidth]{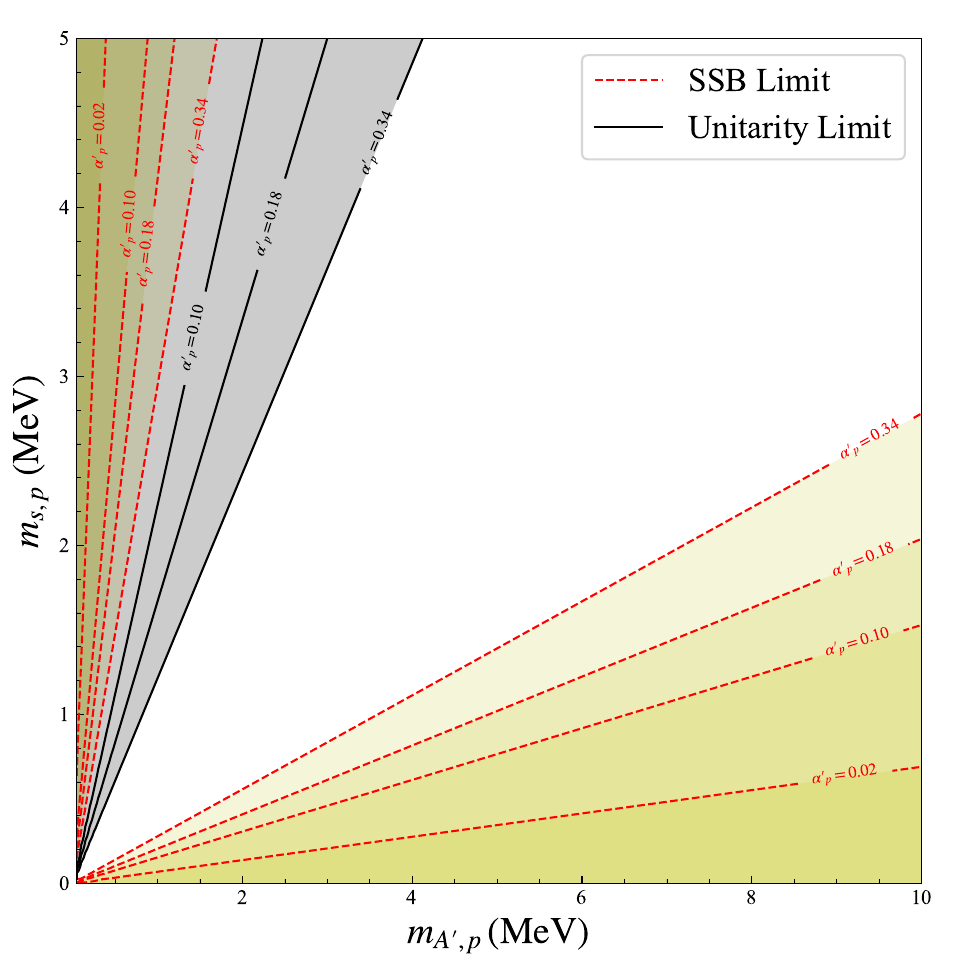}
\caption{The exclusion lines from the SSB condition. The colored regions are excluded. The unitarity limit comes from $\lambda<4\pi$. \textit{Left}: Scale-dependent SSB condition; \textit{Right}: Scale-independent SSB condition.}
\label{fig:MassContour}
\end{figure}

We present Fig.~\ref{fig:MassContour} for illustration. 
From the left panel of Fig.~\ref{fig:MassContour}, it can be seen that the restrictions on the masses largely depend on the renormalization scale $\mu_r$. When $\mu_r$ is fixed, the exclusion lines for different values of $g'$ are intersected on the $m_{A'}$ axis. This is due to  
\begin{equation}
\lim_{m_s\to 0} \Delta V_{\text{1L}}(g',m_s^2,m_{A'}^2,\mu_r)=-\frac{\hbar}{16\pi^2}\times\frac{3}{4}m_{A'}^4\left(\ln\frac{m_{A'}^2}{\mu_r^2}-\frac{5}{6}\right)\,,
\end{equation}
thus the curve $\Delta V_{\text{1L}}=0$ intersects the $m_{A'}$ axis at $m_{A'}=\mu_r e^{5/12}$. 
Such an ambiguous result shows the unreliability of the naive SSB condition $\Delta V_{\text{1L}}(g',m_s^2,m_{A'}^2,\mu_r)>0$.

In the right panel of Fig.~\ref{fig:MassContour}, we used $\Delta V_{p,\text{1L}}(g'_p,m_{s,p}^2,m_{A',p}^2)$, which is independent of the renormalization scale, to constrain the masses. 
We can see that, under a fixed $g_p'$, the curve of $\Delta V_{p,\text{1L}}(g'_p,m_{s,p}^2,m_{A',p}^2)=0$ is a straight line passing through the origin. This is not surprising, because in our model there are only two energy scales $m_{s,p}$ and $m_{A',p}$, so that $\Delta V_{p,\text{1L}}(g'_p,m_{s,p}^2,m_{A',p}^2)$ can be written as 
\begin{equation}
\Delta V_{p,\text{1L}}(g'_p,m_{s,p}^2,m_{A',p}^2)=m_{A',p}^4f(g_p',m_{s,p}^2/m_{A',p}^2)\,.
\end{equation}
with $f(g_p',m_{s,p}^2/m_{A',p}^2)$ being a function induced from Eq.~(\ref{eq:SSBcondition}).
Thus the condition $f(g_p',m_{s,p}^2/m_{A',p}^2)=0$ fixes the mass ratio $m_{s,p}^2/m_{A',p}^2$ with a given $g_p'$.
This is why $\Delta V_{p,\text{1L}}(g'_p,m_{s,p}^2,m_{A',p}^2) = 0$ corresponds to straight lines passing through the origin.

It should be noted that in the right panel of Fig.~\ref{fig:MassContour}, there is also a region in the upper-left cornner that can be excluded by the SSB condition. But this region is covered by the perturbative unitarity bound, and thus the result is actually not reliable.  
Similar situation also happens in the region where $m_{A',p}^2 \ll m_{s,p}^2$ or $m_{A',p}^2 \gg m_{s,p}^2$, because in these regions the expression of $\Delta V_{p,\text{1L}}(g'_p,m_{s,p}^2,m_{A',p}^2)$ suffers from large log terms. 
Resummation method might be required to handle the problem.  
But for the parameter space of our interest in this work, $\log(m_{A',p}^2/m_{s,p}^2)$ is generally not so large, so this problem can be safely ignored. 

\section{Combined constraints on parameter space from data and theoretical requirements \label{sec:fit} }

In this section we use the small-scale data and the theoretical bounds discussed in the preceding two sections to constrain the parameter space  \{$m_\chi$, $m_{A'}$, $\alpha'$, $m_s$\}. 
In the above section, we used \{$m_{A',p}$, $m_{s,p}$, $\alpha'_p$\} to label the pole mass of dark photon, the pole mass of dark Higgs, and the physical coupling strength, respectively. 
While in this section, for the simplicity of notation, we will remove the ``$p$'' subscript and use \{$m_{A'}$, $m_{s}$, $\alpha'$\} . 
Furthermore, in this section the notation ``$v$'' corresponds to the relative velocity of DM inside a halo.  
Strictly speaking, $\alpha'$ is the effective coupling strength at energy scale $m_{A'}$, but this scale is actually not far from the DM momentum transfer inside the halo ($\sim m_\chi v $). 
Therefore, equating $\alpha'$, which describes the interaction strength between DM in the halo, to $\alpha'_p$ defined in the above section,  will not make a big difference.

We will start from the limits on subspace composed by \{$m_\chi$, $m_{A'}$, $\alpha'$\}. 
This subspace is constrained by the requirements on the DM self-scattering cross section from small-scale data. 
We choose the viscosity cross section $\sigma_V\equiv \int d\Omega \sin^2\theta \frac{d\sigma}{d\Omega}$ as the proxy for DM scattering. 
Furthermore, as suggested in~\cite{Colquhoun:2020adl}, we use $\bar{\sigma}= 
\langle \sigma_V v^3 \rangle/24\sqrt{\pi}v^3_0 $ ($v_0$ is the velocity dispersion) as the thermal average cross section because this quantity is more related to energy transfer. 
The public code CLASSICS~\cite{Colquhoun:2020adl} already includes the approximate analytic formula of $\sigma_V$ in different parameter regions and provides a quick calculation of thermal averaging. We will use CLASSICS to calculate $\bar{\sigma}$. 


\begin{table}[htp]
\begin{center}
\begin{tabular}{ c c c }
\hline
\hline
System &\ \ \ $ \langle v \rangle $ (km/s)  \ \ \ &  \ \ \  $\bar{\sigma}/ m_\chi $ (cm$^2$/g)  \ \ \  \\ 
\hline
\hline
MW dSph: UM & \ \ \  30.87 \ \ \  & \ \ \ 40 -- 50 \ \ \  \\
\hline
MW dSph: Draco & \ \ \  56.34 \ \ \  & \ \ \ 20 -- 30 \ \ \  \\ 
\hline
MW dSph: Carina & \ \ \  48.42 \ \ \  & \ \ \ 40 -- 50 \ \ \  \\ 
\hline
MW dSph: Sextans & \ \ \  32.34 \ \ \  & \ \ \ 70 -- 120 \ \ \  \\ 
\hline
MW dSph: CVnI & \ \ \  38.29 \ \ \  & \ \ \ 50 -- 80 \ \ \  \\  
\hline
MW dSph: Sculptor & \ \ \  62.83 \ \ \  & \ \ \ 30 -- 40 \ \ \  \\ 
\hline
MW dSph: Fornax & \ \ \  57.95 \ \ \  & \ \ \ 30 -- 50 \ \ \  \\ 
\hline
MW dSph: LeoII & \ \ \  20.98 \ \ \  & \ \ \ 90 -- 150 \ \ \  \\ 
\hline
MW dSph: LeoI & \ \ \  58.27 \ \ \  & \ \ \ 50 -- 70 \ \ \  \\ 
\hline
UDGs & \ \ \  57.6  \ \ \  & \ \ \ 47 -- 86  \ \ \  \\ 
\hline
Perturber & \ \ \  144 \ \ \  & \ \ \ 14 -- 25 \ \ \  \\ 
\hline
Diversity of dwarfs & \ \ \  $\sim$ 10 \ \ \  & \ \ \ $\sim$ 100 \ \ \  \\ 
\hline
MW-size halos' subhalo evaporation & \ \ \  $\sim$ 200  \ \ \  & \ \ \ < 10  \ \ \  \\ 
\hline
Diversity of galactic rotation curves & \ \ \ $\sim$ 43.2 -- 432  \ \ \  & \ \ \ > 2  \ \ \  \\ 
\hline
Groups & \ \ \  $\sim$ 1152 \ \ \  & \ \ \ < 1 \ \ \  \\ 
\hline
Clusters & \ \ \  $\sim$ 1440 \ \ \  & \ \ \ < 0.1 \ \ \  \\ 
\hline
\end{tabular}
\caption{Small-scale data used in this work to constrain DM scattering.}
\label{data}
\end{center}
\end{table}


Small-scale data put limits on $\bar{\sigma}/m_\chi$. 
To briefly summarize, for an average collision velocity $\langle v \rangle \sim 43.2 - 432 $ km/s, which corresponds to isolated spiral galaxies over a wide mass range, $\bar{\sigma}/m_\chi$ needs to be larger than 2 cm$^{2}$g$^{-1}$ to explain the  diversity of galactic rotation curves~\cite{Kamada:2016euw,Ren:2018jpt,Kaplinghat:2019dhn}. 
For a Milky Way (MW) size halo with $\langle v \rangle \gtrsim 200 $ km/s, $\bar{\sigma}/m_\chi$ should be smaller than 10 cm$^{2}$g$^{-1}$ to avoid the subhalo evaporation~\cite{Vogelsberger:2012ku,Rocha:2012jg,Zavala:2012us}.
For Milky Way’s dwarf spheroidal galaxies (MW dSphs) with $\langle v \rangle \sim 20 - 60 $ km/s,  $\bar{\sigma}/m_\chi \sim \mathcal{O}(10) - \mathcal{O}(100) $  cm$^{2}$g$^{-1}$ is required to explain the observed anti-correlation between the central DM densities and orbital pericenter distances~\cite{Correa:2020qam}. 
To explain the diversity of dwarf galaxies within and surrounding the MW, 
$\bar{\sigma}/m_\chi$ should be around 100 cm$^{2}$g$^{-1}$ when $\langle v \rangle \sim 10 $ km/s~\cite{Yang:2022mxl}. 
To explain the extremely dense DM substructure that perturbs the strong lens galaxy SDSSJ0946+1006~\cite{Minor:2020hic} (this observation will be labeled as `perturber'), 
$\bar{\sigma}/m_\chi$ needs to be approximately $14-25$ cm$^{2}$g$^{-1}$ when $\langle v \rangle \sim 144 $ km/s~\cite{Nadler:2023nrd}.
To explain the rotation curves of isolated and gas-rich ultradiffuse galaxies (UDGs), $\bar{\sigma}/m_\chi$ should be around $47-86$ cm$^{2}$g$^{-1}$ when $\langle v \rangle \sim 57.6 $ km/s~\cite{Nadler:2023nrd}~\footnote{For perturber and UDGs, Ref.~\cite{Nadler:2023nrd} only provides the upper limit. The lower limit is obtained via the discussion with the author of Ref.~\cite{Nadler:2023nrd}.}.  
Finally, for groups and clusters with $\langle v \rangle$ around 1152 km/s and 1440 km/s, $\bar{\sigma}/m_\chi$ should be smaller than 1 cm$^{2}$g$^{-1}$ and 0.1 cm$^{2}$g$^{-1}$, respectively~\cite{Sagunski:2020spe,Kaplinghat:2015aga,Andrade:2020lqq}. 
These data are summarized in Table ~\ref{data}.

\begin{figure}[ht]
\centering
\includegraphics[width=7.0in]{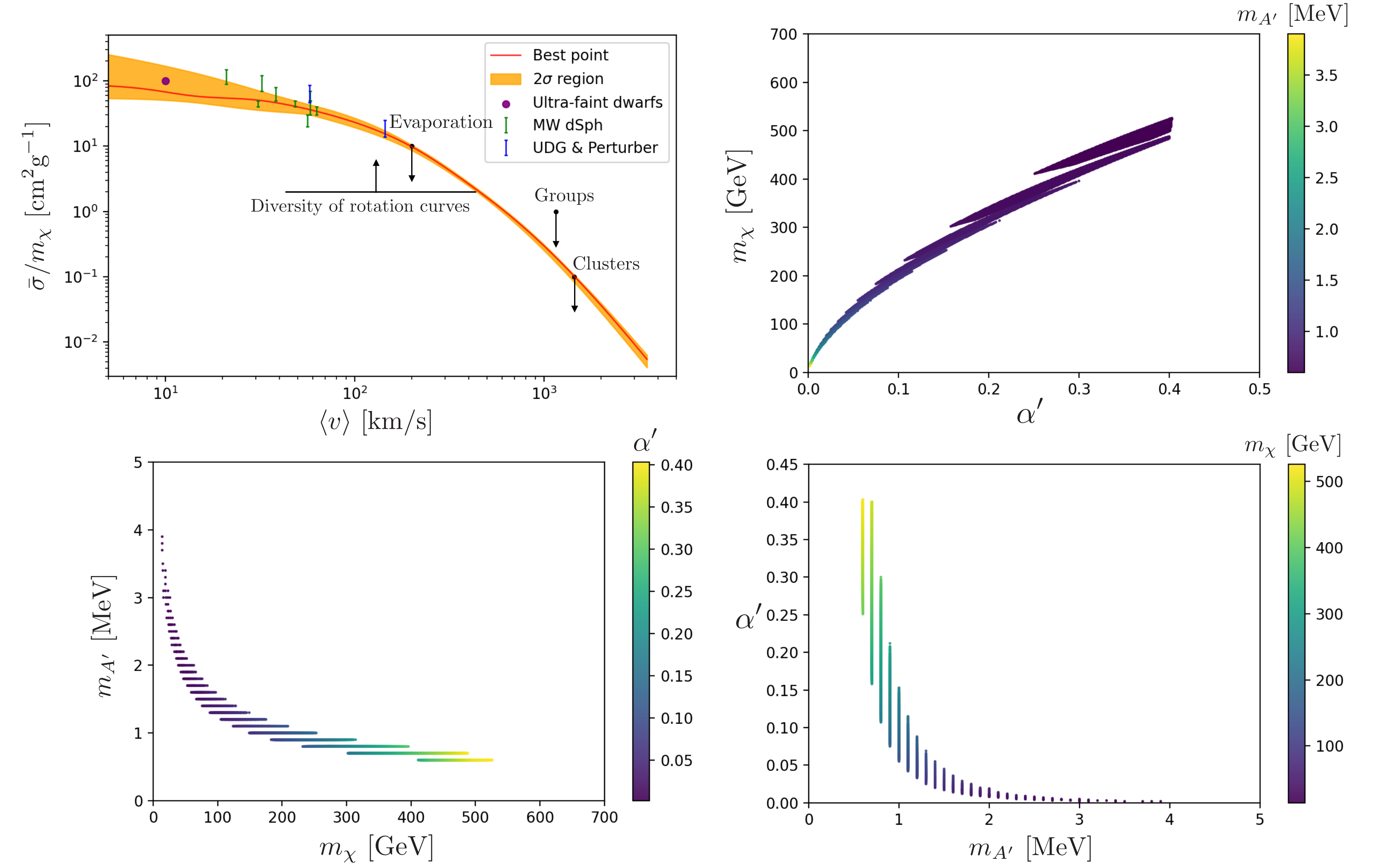}
\caption{\textit{Upperleft}: the thermal average cross-section per unit mass $\bar{\sigma}/m_\chi$ as a function of the average collision velocity $\langle v \rangle$. The red line comes from the best fit point, and the orange region corresponds to the 2$\sigma$ region. The data presented and used in the plot are summarized in Table ~\ref{data}.  
\textit{Upperright}: the 2$\sigma$ parameter space projected on $\alpha'$-$m_\chi$ plane, as color mapped by the value of $m_{A'}$.
\textit{Lowerleft}: the 2$\sigma$ parameter space projected on $m_\chi$-$m_{A'}$ plane, as color mapped by the value of $\alpha'$.
\textit{Lowerright}: the 2$\sigma$ parameter space projected on $m_{A'}$-$\alpha'$ plane, as color mapped by the value of $m_\chi$.  }
\label{fig:fit}
\end{figure}

We perform a $\chi^2$ fit to find the 2$\sigma$ region in the parameter space that is favored by small-scale data. 
In the $\chi^2$ value calculation we only use data from MW dSphs, UDGs, and perturber. 
Other data, except diversity of dwarfs, are treated as bounds. 
For the diversity of dwarfs, Ref.~\cite{Yang:2022mxl} only provides one benchmark parameter point instead of a parameter range to explain the observation, thus we will not use this one single point in the fit. 
The ranges of parameters in our study are 
\begin{eqnarray}
m_\chi < \text{1 TeV} \ , \ m_{A'} < \text{10 MeV} \ , \ \alpha' < 0.5
\end{eqnarray}
The coupling strength $\alpha'$ cannot be too large, otherwise the Landau pole will appear at a quite low scale. 
The one-loop beta function of $\alpha'$ in our model is 
\begin{eqnarray}
\frac{d \alpha}{d \ln \mu} =  
\left\{
\begin{array}{lr}
   \frac{2}{3\pi}\frac{1}{4}\alpha'^2 ,  & \mu < m_\chi \\
   \frac{2}{3\pi}\frac{5}{4}\alpha'^2 ,  & \mu \geq m_\chi
\end{array}
\right.
\end{eqnarray}
Thus $\alpha'$ always increases as energy scale increases.
Taking into account the effectiveness of perturbation, we further require $\alpha'$ to be smaller than 1.0 below 10 TeV. 

In Fig.~\ref{fig:fit} we present our fit results. 
The upper-left panel of Fig.~\ref{fig:fit} shows $\bar{\sigma}/m_\chi$ as a function of $\langle v \rangle$. 
For an intuitive display, all the data in Table~\ref{data} are also present by bars, arrows, or spot. 
The curves of the best fit point and the 2$\sigma$ region are given. 
The other parts of Fig.~\ref{fig:fit} show the 2$\sigma$ parameter space projected on different two-dimensional parameter planes. 
The positive correlation between $\alpha'$ and $m_\chi$ shown in the upperright of Fig.~\ref{fig:fit} simply indicate that $\alpha'^2$ and DM number density (which gets smaller as $m_\chi$ grows) are roughly inversely proportional under a certain DM scattering probability. 
The lower-left panel of Fig.~\ref{fig:fit} can also be understood via a similar argument. 
When DM velocity $v$ become very small (e.g. inside an ultra-faint dwarf), the crosssection is roughly proportional to $(m^2_{A'})^{-2}$ (square of t-channel propagator), and this why there is a negative correlation between $m_\chi$ and $m_{A'}$. 

\begin{figure}[ht]
\centering
\includegraphics[width=7.0in]{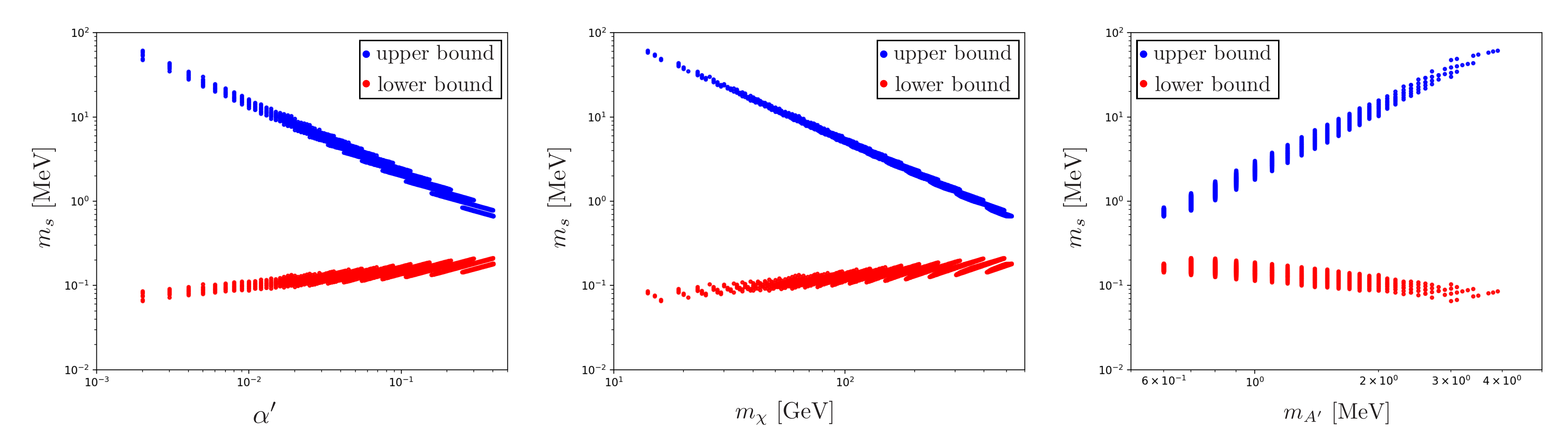}
\caption{ The upper and lower bounds on the dark Higgs mass $m_s$ as functions of $\alpha'$, $m_\chi$, and $m_{A'}$, respectively. The blue and red dots correspond to upper and lower bounds, respectively. 
The upper bound comes from the tree-level perturbative unitary limit in Sec.~\ref{sec:upper}. 
The lower bound comes from the one-loop level Linde-Weinberg limit in Sec.~\ref{sec:lower}. }
\label{fig:massbound}
\end{figure}

Next we use the theoretical limits discussed in the preceding two sections to further constrain the range of $m_s$. 
The upper bound of $m_s$ comes from the tree-level perturbative unitarity requirement discussed in Sec.~\ref{sec:upper}, which is quite simple (see Eq.~(\ref{massuplimit})): 
\begin{eqnarray}
m_s^2\le 2\pi v_{\rm cl}^2    \  \ \Rightarrow \ \ m_s^2\le  \frac{m^2_{A'}}{2 \alpha'} 
\end{eqnarray}
The tree-level mass-parameter relation, i.e. Eq.~(\ref{mass_relation}), is used here since we are considering the tree-level upper bound. 
On the other hand, the one-loop Linde-Weinberg lower bound, which is gauge and scale independent, is determined by solving the SSB condition: 
\begin{eqnarray}
\Delta V_{p,\text{1L}}(g',m_{s}^2,m_{A'}^2)>0 
\end{eqnarray}
with the expression of $\Delta V_{p,\text{1L}}$ given in Eq.~({\ref{eq:SSBcondition}}). 

In Fig.~\ref{fig:massbound} we present the upper and lower bounds on the dark Higgs mass $m_s$ as functions of $\alpha'$, $m_\chi$, and $m_{A'}$, respectively. 
It can be seen that the lower bound of $m_s$ in our SIDM model is not sensitive to the parameters, always locating around $\mathcal{O}(0.1)$ MeV.  
As contrast, the upper bound of $m_s$ can differ by two orders of magnitude. 
This is simply because the allowed range of $\alpha'$ is relatively large.
But in the entire allowed parameter space, the dark Higgs cannot be heavier than $\mathcal{O}(100)$ MeV. 
Therefore, in the phenomenological study of the dark sector, generally speaking we cannot simply ignore the dark Higgs. 
The specific phenomenological impact of such a light dark Higgs is beyond the scope of this paper, which will be studied in our future work. 

\section{Conclusion and discussion \label{sec:conclu} }

Motivated by the null results of current DM searches and the cold dark matter small-scale problems, in this work we studied a dark sector charged by a spontaneous broken gauge $U(1)'$.  
This dark sector model can be fully described by the dark matter mass $m_\chi$, the dark mediator (dark photon) mass $m_{A'}$, the dark coupling strength (dark fine-structure constant) $\alpha'\equiv \frac{g'^2}{4\pi}$, and the dark Higgs mass $m_s$. 
Previous studies focused on how to constrain the three-dimensional parameter subspace $\{m_\chi,\ m_{A'}, \ \alpha'  \}$ by using the small-scale data. 
In this work we extended the study to constrain the four-dimensional parameter space $\{m_\chi,\ m_{A'}, \ \alpha', \ m_s  \}$ by further considering the theoretical bounds on the dark Higgs mass. 
For the upper bound on $m_s$ we considered the tree-level perturbative unitarity, while for the lower bound 
we considered the one-loop Linde-Weinberg bound. 
Combining the theoretical and observational constraints, we obtained the following constraints:
\begin{eqnarray}
 10 \text{ GeV} \lesssim  m_\chi \lesssim 500 \text{ GeV} \ , \  0.5 \text{ MeV} \lesssim m_{A'} \lesssim 5 \text{ MeV} \ , \\\nonumber
 \  0.001 \lesssim \alpha'  \lesssim 0.4 \ , \   0.05 \text{ MeV}  \lesssim m_s  \lesssim  50 \text{ MeV} 
\end{eqnarray}

Finally, we give some remarks:
\begin{itemize}
\item The one-loop Linde-Weinberg lower bound obtained in this work is quite different from others as it is independent of gauge and energy scale, thus is more physically reliable.
\item In reality, the SSB of dark $U(1)'$ happens in the expanding Universe. Thus, to make the SSB really happen, we need to compare the nucleation rate with the expansion rate, and this may impose a stricter limit than Linde-Weinberg bound~\cite{Guth:1980zk}. But this bound depends on the temperature of the dark sector, provided that the phase transition is dominated by thermal fluctuation. 
\item Our fitting results indicated that the dark Higgs in our model cannot be ignored in the phenomenological study of dark sector.  We will investigate it in our future work. For some related studies,  see, e.g., ~\cite{Kim:2016fdv,Bell:2016uhg}.
\end{itemize}

\addcontentsline{toc}{section}{Acknowledgments}
\acknowledgments
M.Z. appreciates helpful discussions with Daneng Yang. This work was supported by the Natural Science Foundation of China (NSFC) under grant numbers 12105118, 11947118, 12075300, 11821505 and 12335005,  by the Research Fund for Outstanding Talents (5101029470335) from Henan Normal University, by the Peng-Huan-Wu Theoretical Physics Innovation Center (12047503) funded by the National Natural Science Foundation of China, by the CAS Center for Excellence in Particle Physics (CCEPP), and by the Key Research Program of the Chinese Academy of Sciences under grant No. XDPB15.
\vspace{0.5cm}

\appendix
\section{Derivation of unitarity bound} \label{appA}

In this appendix we provide a detailed derivation of the unitary bound that one can easily follow and check. 
The notation used here is slightly different from that used in the main text.

\subsection{Unitarity constraints on subspaces\label{subsec:UnitarityOnSubspace} }

\begin{figure}[ht]
\centering
\includegraphics[width=3in]{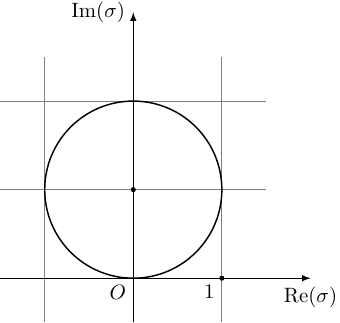}
\caption{The area inside the circle is allowed by unitarity.}
\label{fig:UnitarityConstraint}
\end{figure}

Consider an incoming state $\ket{a}$ which is normalized. 
Due to conservation of probability, we have
\begin{eqnarray}
1=\sum_{n\in I}\abs{\mel{n}{S}{a}}^2=\sum_{n\in I}\mel*{a}{S^\dagger}{n}\mel{n}{S}{a}=\mel{a}{S^\dagger S}{a} \, ,
\end{eqnarray}
where $I$ is the index set of a complete orthogonal basis. $I$ may contain continuous parts, in which case the summation in the above equation needs to be replaced with integral. The above equation implies $S^\dagger S=\mathbf{1}$, which is the well-known unitarity. In practical calculations, we generally cannot start from the entire Hilbert space $\mathcal{H}$ to verify whether unitarity holds.
Instead, we only consider a subspace $\mathcal{H}'$ of $\mathcal{H}$, and the index set of the corresponding orthogonal basis is $I'\subset I$. In this case, we have
\begin{eqnarray}
1\ge\sum_{n\in I'}\abs{\mel{n}{S}{a}}^2=\sum_{n\in I'}\mel*{a}{S^\dagger}{n}\mel{n}{S}{a} \, .
\end{eqnarray}
If we denote the scattering matrix on this subspace as $S'$, then the above equation implies $S'^\dagger S'\le \mathbf{1}_{\mathcal{H}'}$ (the meaning of the relation $a\le b$ between Hermitian operators is that $b-a$ is positive definite). In practical calculations, what we first obtain is generally the scattering amplitude $M'$, which is related to $S'$ by $S'=\mathbf{1}_{\mathcal{H}'}+i M'$. If $M'$ has an eigenvalue $\sigma$ and the corresponding eigenstate is $\ket{\sigma}$, then
\begin{eqnarray}
\abs{1+i\sigma}^2=\mel*{\sigma}{S'^\dagger S}{\sigma}\le 1  \, .
\end{eqnarray}
In other words, $1+i\sigma$ can only fall within a circle centered at the origin with a radius of 1. Equivalently, $\sigma$ can only fall within a circle with center of $i$ and radius of $1$, as shown in Fig.~\ref{fig:UnitarityConstraint}. 

The matrix $M'$ is generally unitarily diagonalizable, which means that for any normalized state $\ket{a}$, the expectation value $\mel*{a}{M'}{a}$ will fall within a circle with center of $i$ and radius of $1$. In the calculation of perturbative unitarity, we generally calculate to the tree-level, so the obtained matrix $M'$ is a real symmetric matrix, and the corresponding eigenvalue $\sigma$ is a real number.
This result  seems to violate the requirement of unitarity unless $\sigma=0$. However, as we can see from Fig.~\ref{fig:UnitarityConstraint}, the unitarity boundary of $\sigma$ is tangent to the $\Re(\sigma)$ axis at the origin. 
So even $\sigma$ is a real number at tree-level, 
unitarity will not be violated if the loop-level corrections provide $\sigma$ with positive imaginary part. 
Based on these arguments, we should use  following conditions as unitarity constraints at  tree-level:
\begin{eqnarray}
    \abs{\sigma}\le 2\,,\quad\quad \abs{\Re(\sigma)}\le 1  \, .
\end{eqnarray}

Since our goal is to calculate the eigenvalues of $M$ (from now on, we use $M$ to refer to scattering amplitude in subspace $\mathcal{H}'$), it would be much more convenient to first diagonalize $M$ with some conserved quantities and then analyze it. In a quantum field theory which is invariant under Poincaré transformation, both four-momentum and angular momentum are conserved quantities. But the total four-momentum operator and the total angular momentum operator do not commute, so we cannot decompose $M$ according to their common invariant subspaces. However, as we will see later, the motion of the system can be decomposed into global motion and relative motion, and the operators of the global motion part commute with the operators of the relative motion part, so we can take a complete set of common eigenstates of the total four-momentum and relative angular momentum. Decomposing $M$ according to the common invariant subspaces of the total four-momentum $\hat{\mathbf{p}}^\mu$ and the relative angular momentum $\{J^2,J_z\}$, we get
\begin{eqnarray}
M=\bigoplus_{p,j,m}M_{p}^{j,m}   \, .
\end{eqnarray}
Under this decomposition, we only need to calculate the eigenvalue of $M$ under specific total momentum $p^\mu$ and relative angular momentum $\{j,m\}$.
We will consider the two-particle states among them, which is the subspace $\mathcal{H}'$ we take.

\subsection{Decomposition of total motion and relative motion for two-particle state \label{subsec:decomposition} }

In quantum field theory, the methods to calculate amplitudes are all for the cases where both incoming and outgoing particles have specific momentum, where the two-particle state can be written as $\ket{\mathbf{p}_1,\mathbf{p}_2;\eta}$, with $\eta$ representing some internal quantum numbers of the particle. In the following we will not explicitly write out $\eta$ for brevity.

According to the discussion in the previous subsection, we need to decompose the two-particle subspace of the incoming and outgoing state into the common invariant subspaces of the total four-momentum $\hat{\mathbf{p}}^\mu$ and the relative angular momentum $\{J^2,J_z\}$. To achieve this, we can proceed in two steps: firstly, we decompose the whole state into total motion and relative motion parts; secondly, we expand the relative motion part according to the angular momentum, which is the partial wave expansion.

We start with the first step. In non-relativistic quantum mechanics, there exists a position operator, so for two particles, their motion can be easily decomposed into relative and total parts. Since there is no well-behaved position operator in quantum field theory~\cite{Newton:1949cq,kaiser2018lectures}, we cannot define the center-of-mass as in classical mechanics, and thus we cannot simply decompose the motion into total motion and relative motion. However, in quantum field theory we can decompose the Hilbert space by Wigner's little group representation~\cite{Wigner:1939cj,Bargmann:1948ck,weinberg2005quantum}:
\begin{eqnarray}
\mathcal{H}=\mathcal{H}_f\otimes \mathcal{H}_\mathrm{r}  \, .
\end{eqnarray}
Under this decomposition, a momentum eigenstate can be written as $\ket{p}\otimes\ket{\alpha}$, satisfying $\hat{\mathbf{p}}^\mu(\ket{p}\otimes\ket{\alpha})=p^\mu (\ket{p}\otimes\ket{\alpha})$. The state $\ket{\alpha}$ is translation invariant. Under Lorentz transformations, $\ket{\alpha}$ only transforms according to the spatial rotation part (for massive states). This kind of decomposition actually corresponds to the decomposition of total motion and relative motion in classical mechanics, where $\ket{p}$ represents total motion and $\ket{\alpha}$ represents "relative motion". Applying this decomposition to the two-particle state $\ket{\mathbf{p}_1,\mathbf{p}_2}$, we get
\begin{equation}
\ket{\mathbf{p}_1,\mathbf{p}_2}=\ket{p_\mathrm{t}}\otimes\ket{\mathbf{p}_\mathrm{r}}\,,\label{eq:total-ref-decomposition}
\end{equation}
where $p_\mathrm{t}=(p^0_\mathrm{t},\mathbf{p}_\mathrm{t})=(p_1^0+p_2^0,\mathbf{p}_1+\mathbf{p}_2)$, and $\mathbf{p}_\mathrm{r}$ is the relative momentum of the two particles. We do not have a simple expression for $\mathbf{p}_\mathrm{r}$ from $\mathbf{p}_1$ and $\mathbf{p}_2$, but we do not need such an expression. In principle, we should first find a boost $\Lambda$ such that $\Lambda p_\mathrm{t}=(E_{\rm cm},0)$, then $\mathbf{p}_\mathrm{r}$ can be obtained via $\Lambda p_1=(\sqrt{m_1^2+p_\mathrm{r}^2},\mathbf{p}_\mathrm{r})$. 
The symbols $E_{\rm cm}$ used here is the total energy in the center of mass frame, and $p_\mathrm{r}=\abs{\mathbf{p}_\mathrm{r}}$ is the magnitude of the relative momentum.

Firstly we need to deal with the normalization of $\ket{p_\mathrm{t}}\otimes\ket{\mathbf{p}_\mathrm{r}}$. 
For non-identical particles, the normalization condition for the state $\ket{\mathbf{p}_1,\mathbf{p}_2}$ is
\begin{equation}
\braket{\mathbf{p}'_1,\mathbf{p}'_2}{\mathbf{p}_1,\mathbf{p}_2}=(2\pi)^6 4p_1^0p_2^0\delta^3(\mathbf{p}_1-\mathbf{p}'_1)\delta^3(\mathbf{p}_2-\mathbf{p}'_2)\,.
\end{equation}
The total four-momentum does not need to satisfy on-shell condition, so $\braket{p'_\mathrm{t}}{p_\mathrm{t}}=(2\pi)^4\delta^4(p_\mathrm{t}-p'_\mathrm{t})$. According to Eq.~\eqref{eq:total-ref-decomposition}, we have
\begin{equation}
(2\pi)^4\delta^4(p_{\rm{t}}-p'_{\rm{t}})\braket{\mathbf{p}'_\mathrm{r}}{\mathbf{p}_\mathrm{r}}=(2\pi)^6 4p_1^0p_2^0\delta^3(\mathbf{p}_1-\mathbf{p}'_1)\delta^3(\mathbf{p}_2-\mathbf{p}'_2)\, .\label{eq:nomalization01}
\end{equation}
If we define
\begin{equation}
\mathcal{N}\equiv\abs{\frac{\partial^6(\mathbf{p}_\mathrm{t},\mathbf{p}_\mathrm{r})}{\partial^6(\mathbf{p}_1,\mathbf{p}_2)}}\, ,
\end{equation}
then we have
\begin{align}
\delta^3(\mathbf{p}_1-\mathbf{p}'_1)\delta^3(\mathbf{p}_2-\mathbf{p}'_2) &= \frac1{\mathcal{N}}\delta^3(\mathbf{p}_\mathrm{t}-\mathbf{p}'_\mathrm{t})\delta^3(\mathbf{p}_\mathrm{r}-\mathbf{p}'_\mathrm{r})\nonumber \\
& = \frac1{\mathcal{N}}\delta^3(\mathbf{p}_\mathrm{t}-\mathbf{p}'_\mathrm{t}) \frac1{p_\mathrm{r}^2}\delta(p_\mathrm{r}-p_\mathrm{r}')\delta(\Omega-\Omega')\,,
\end{align}
where $\delta(\Omega-\Omega')$ is defined by $\delta(\Omega-\Omega') \equiv \delta(\cos\theta-\cos\theta')\delta(\phi-\phi')$, with $\Omega=(\theta,\phi)$ being the polar angle of $\mathbf{p}_\mathrm{r}$. The expression of $\mathcal{N}$ is quite complex, but fortunately, it can be shown that
\begin{equation}
\mathcal{N}\Big|_{\mathbf{p}_\mathrm{t}=0}=\abs{\frac{\partial^6(\mathbf{p}_\mathrm{t},\mathbf{p}_\mathrm{r})}{\partial^6(\mathbf{p}_1,\mathbf{p}_2)}}_{\mathbf{p}_\mathrm{t}=0}=1\, .
\end{equation}
Therefore, in the center-of-mass frame we have a quite simple expression
\begin{equation}
\delta^3(\mathbf{p}_1-\mathbf{p}'_1)\delta^3(\mathbf{p}_2-\mathbf{p}'_2) = \delta^3(\mathbf{p}_\mathrm{t}-\mathbf{p}'_\mathrm{t}) \frac1{p_\mathrm{r}^2}\delta(p_\mathrm{r}-p_\mathrm{r}')\delta(\Omega-\Omega')\, .
\end{equation}

In the center of mass frame,  $p^0_\mathrm{t}=\sqrt{m_1^2+p^2_\mathrm{r}}+\sqrt{m_2^2+p^2_\mathrm{r}}$, and thus
\begin{equation}
\dd{p^0_\mathrm{t}}=\frac{p_\mathrm{r}\dd{p_\mathrm{r}}}{p_1^0}+\frac{p_\mathrm{r}\dd{p_\mathrm{r}}}{p_2^0}=\frac{p_\mathrm{r} E_{\rm cm}\dd{p_\mathrm{r}}}{p_1^0 p_2^0} \, .
\end{equation}
So we have
\begin{equation}
\delta(p_\mathrm{r}-p'_\mathrm{r})=\frac{p_\mathrm{r} E_{\rm cm}}{p_1^0 p_2^0}\delta(p^0_\mathrm{t}-p'^0_\mathrm{t}) \,.
\end{equation}
Therefore, 
\begin{equation}
\delta^3(\mathbf{p}_1-\mathbf{p}'_1)\delta^3(\mathbf{p}_2-\mathbf{p}'_2) = \frac{1}{p_1^0 p_2^0}\frac{E_{\rm cm}}{p_\mathrm{r}}\delta^4(p_\mathrm{t}-p_\mathrm{t}')\delta(\Omega-\Omega') \, .\label{eq:DeltaFunctionRelation}
\end{equation}
Comparing Eq.~\eqref{eq:DeltaFunctionRelation} with Eq.~\eqref{eq:nomalization01}, we get
\begin{equation}
\braket{\mathbf{p}'_\mathrm{r}}{\mathbf{p}_\mathrm{r}}=(2\pi)^2\frac{4E_{\rm cm}}{p_\mathrm{r}}\delta(\Omega-\Omega') \,.
\end{equation}

Although this result is obtained in the center-of-mass frame, it also applies to the general case because the inner product of states is Lorentz invariant. If you want to use the above equation in a general reference frame, we can just take the quantities on the right-hand side as the quantities in the center-of-mass frame.

We can use the polar angles of $\mathbf{p}_r$ to label $\ket{\mathbf{p}_\mathrm{r}}$ and define
\begin{equation}
\ket{\theta,\phi}\equiv\frac{1}{2\pi}\sqrt{\frac{p_\mathrm{r}}{4E_{\rm cm}}}\ket{\mathbf{p}_\mathrm{r}}\, ,
\end{equation}
so that
\begin{equation}
\braket{\theta',\phi'}{\theta,\phi}=\delta(\Omega-\Omega')\, .
\end{equation}

Due to the conservation of total momentum, the scattering amplitude $M$ can be decomposed by $M=\mathbf{1}\otimes M_{p_\mathrm{t}}$, and thus
\begin{equation}
i\mel*{\mathbf{p}'_1,\mathbf{p}'_2}{M}{\mathbf{p}_1,\mathbf{p}_2}=i (2\pi)^4\delta^4(p_{\rm{t}}-p'_{\rm{t}})\mel*{\mathbf{p}'_\mathrm{r}}{M_{p_\mathrm{t}}}{\mathbf{p}_\mathrm{r}}\,.
\end{equation}

From this, we can see that the scattering amplitude that we usually calculate by using Feynman diagrams and Feynman rules is actually $\mel*{\mathbf{p}'_\mathrm{r}}{M_{p_\mathrm{t}}}{\mathbf{p}_\mathrm{r}}$. This point is very important and directly related to the quantities we should use to measure the violation of unitarity. 
For later convenience, we can also denote scattering amplitude by polar angles
\begin{equation}
\begin{aligned}
M((\theta,\phi)\to(\theta',\phi')) &= \mel*{\mathbf{p}'_\mathrm{r}}{M_{p_\mathrm{t}}}{\mathbf{p}_\mathrm{r}}\\
&= (2\pi)^2\frac{4E_{\rm cm}}{\sqrt{p_\mathrm{r} p'_\mathrm{r}}}\mel*{\theta',\phi'}{M_{p_\mathrm{t}}}{\theta,\phi}\,.
\end{aligned}\label{eq:ampt}
\end{equation}

\subsection{Helicity and partial wave expansion of state \texorpdfstring{$\ket{\theta,\phi}$}{|theta,phi>}\label{subsec:PWEofState} }

After completing the first step, we need to expand $\ket{\theta,\phi}$ according to the angular momentum. The relative motion part has three angular momentum: $\mathbf{L}$, $\mathbf{S}_1$ and $\mathbf{S}_2$, where $\mathbf{L}$ is the orbital angular momentum of relative motion. 
The choice of complete set of commutative observables (CSCOs) is somewhat arbitrary. 
For example, we can choose $\{L^2,L_z,S_1^2,S_{1z},S_2^2,S_{2z}\}$ or $\{J^2,J_z,L^2,S^2,S_1^2,S_2^2\}$ as our CSCOs. 
For our convenience, it is better to choose the following CSCOs
\begin{eqnarray}
    \{J^2,J_z,S_1^2,S_2^2,\Lambda_1,\Lambda_2\}
\end{eqnarray}
 where $\Lambda_{1,2}=\mathbf{S}_{1,2}\cdot\hat{\mathbf{p}}_{\rm r}$ are helicities of particle 1 and 2 in the 
 center-of-mass frame, and $\hat{\mathbf{p}}_\mathrm{r}=\mathbf{p}_\mathrm{r}/p_\mathrm{r}$ is the direction of relative motion.
 In fact, the total helicity $\Lambda=\mathbf{J}\cdot\hat{\mathbf{p}}_{\rm r}$ (do not confuse this symbol with Lorentz transformation) also commutes with the above set of operators. 
 But $\Lambda$ is not independent:
\begin{eqnarray}
\Lambda=\mathbf{J}\cdot\hat{\mathbf{p}}_\mathrm{r}=\mathbf{S}_1\cdot\hat{\mathbf{p}}_\mathrm{r}+\mathbf{S}_2\cdot\hat{\mathbf{p}}_\mathrm{r}=\mathbf{S}_1\cdot\hat{\mathbf{p}}_\mathrm{r}-\mathbf{S}_2\cdot(-\hat{\mathbf{p}}_\mathrm{r})=\Lambda_1-\Lambda_2\,
\end{eqnarray}
where $\mathbf{L}\cdot\hat{\mathbf{p}}_\mathrm{r}=0$ is used. From the above equation, we know that the total helicity of the relative motion part is equal to the difference of the helicities of each particle. Particularly, in the 
center-of-mass frame, the helicity $\mathbf{J}_i\cdot\hat{\mathbf{p}}_i$ defined by each particle in the single particle form is exactly equal to its relative motion part of the helicity, because
\begin{eqnarray}
\mathbf{J}_1\cdot\hat{\mathbf{p}}_1=\mathbf{L}_1\cdot\hat{\mathbf{p}}_1+\mathbf{S}_1\cdot\hat{\mathbf{p}}_1=\mathbf{S}_1\cdot\hat{\mathbf{p}}_\mathrm{r}=\Lambda_1 \, .
\end{eqnarray}
Similar derivation can be done for $\mathbf{J}_2\cdot\hat{\mathbf{p}}_2$. If it is not in the center-of-mass frame, this relationship is not so simple.

After considering the helicity, the equations in the previous section does not need to be changed much, we only need to add helicity index to the state. 
In the center-of-mass frame, a two-particle state with both particles moving along the $z$ axis can be written as $\ket{\theta=0,\phi=0,\lambda_1,\lambda_2}$, or $\ket{0,0,\lambda_1,\lambda_2}$ in short. 
Here $\lambda_1$ and $\lambda_2$ are the helicities of the two particles, and the total helicity is $\lambda=\lambda_1-\lambda_2$.
Since the relative momentum of the two particles points to the positive direction of the $z$ axis, the $z$ component of the total angular momentum is equal to the total helicity $\lambda$. Therefore, the partial wave expansion of $\ket{0,0,\lambda_1,\lambda_2}$ can be written as
\begin{eqnarray}
\ket{0,0,\lambda_1,\lambda_2}=\sum_{j=\abs{\lambda}}^\infty C_j \ket{j,\lambda;\lambda_1,\lambda_2}\,.
\end{eqnarray}
The right-hand side of the above equation uses $\{J^2,J_z,S_1^2,S_2^2,\Lambda_1,\Lambda_2\}$ as the CSCOs, and $C_j$ is the coefficient to be determined. Perform a spatial rotation to make the relative momentum point to the $(\theta,\phi)$ direction, then
\begin{eqnarray}
\ket{\theta,\phi,\lambda_1,\lambda_2}=\sum_{j=\abs{\lambda}}^\infty\sum_{m=-j}^j C_j D^j_{m\lambda}(\phi,\theta,0)\ket{j,m;\lambda_1,\lambda_2}\,,
\end{eqnarray}
where $D^j_{m\lambda}$ is the matrix element of the $(2j+1)$-dimensional irreducible representation of the rotation group. Some literature also use $D^j_{m\lambda}(\phi,\theta,-\phi)$, which will not affect the final result. In the process of obtaining the above equation, we used the rotation invariance of the helicity. In order to obtain the coefficient $C_j$, we take the inner product of the states based on the above equation to get
\begin{eqnarray}
\delta(\cos\theta-\cos\theta')\delta(\phi-\phi') 
&=&  \braket{\theta',\phi',\lambda_1,\lambda_2}{\theta,\phi,\lambda_1,\lambda_2}\nonumber\\
&=&  \sum_{j',m'}\sum_{j,m}C_{j'}^*C_jD^{j'*}_{m'\lambda}(\phi',\theta',0)D^j_{m\lambda}(\phi,\theta,0)\braket{j',m';\lambda_1,\lambda_2}{j,m;\lambda_1,\lambda_2}\nonumber\\
&=&  \sum_{j,m}\abs{C_j}^2D^{j*}_{m\lambda}(\phi',\theta',0)D^j_{m\lambda}(\phi,\theta,0)\, .
\end{eqnarray}
By using the relationship $D^j_{m\lambda}(\phi,\theta,0)=e^{-i\phi m}d^j_{m\lambda}(\theta)$, we can get
\begin{eqnarray}
\sum_{j,m}\abs{C_j}^2 e^{-i(\phi-\phi')m}d^j_{m\lambda}(\theta')d^j_{m\lambda}(\theta)=\delta(\cos\theta-\cos\theta')\delta(\phi-\phi')\,.
\end{eqnarray}
Multiplying both sides of the above equation by $e^{-i m_1\phi'}$ and then integrating $\phi'$ from $0$ to $2\pi$, we obtain
\begin{eqnarray}
2\pi\sum_{j=\max(\abs{m},\abs{\lambda})}^\infty\abs{C_j}^2d^j_{m\lambda}(\theta')d^j_{m\lambda}(\theta)=\delta(\cos\theta-\cos\theta')\,,
\end{eqnarray}
where we replaced $m_1$ with $m$. Multiplying both sides of the above equation by $d^{j_1}_{m\lambda}(\theta')\sin\theta'\dd\theta'$ and integrating by using the orthogonality relation of the $d$-functions
\begin{eqnarray}
\int_{0}^\pi d^j_{m\lambda}(\theta')d^{j_1}_{m\lambda}(\theta')\sin\theta'\dd\theta'=\frac{2}{2j+1}\delta_{jj_1}\, 
\end{eqnarray}
we can get
\begin{eqnarray}
\abs{C_j}^2=\frac{2j+1}{4\pi}\,.
\end{eqnarray}
Without loss of generality, we can take $C_j$ to be
\begin{eqnarray}
C_j=\sqrt{\frac{2j+1}{4\pi}}\,.
\end{eqnarray}
Therefore, we obtain
\begin{eqnarray}
\ket{\theta,\phi,\lambda_1,\lambda_2}=\sum_{j=\abs{\lambda}}^\infty\sum_{m=-j}^j \sqrt{\frac{2j+1}{4\pi}} D^j_{m\lambda}(\phi,\theta,0)\ket{j,m;\lambda_1,\lambda_2}\,.
\end{eqnarray}
The inverse transformation is
\begin{equation}
\ket{j,m;\lambda_1,\lambda_2}=\sqrt{\frac{2j+1}{4\pi}}\int \dd{\Omega}D^{j*}_{m\lambda}(\phi,\theta,0)\ket{\theta,\phi,\lambda_1,\lambda_2}\,.\label{eq:inverse-partial-wave-exp}
\end{equation}

\subsection{Partial wave expansion of the amplitude \label{subsec:PWEofAMP} }

Assuming the momenta of the incoming particles are parallel to the $z$-axis, then according to Eq.~\eqref{eq:ampt} we have
\begin{align}
M(\theta,\phi) &= (2\pi)^2\frac{4E_{\rm cm}}{\sqrt{p_\mathrm{r} p'_\mathrm{r}}}\mel*{\theta,\phi,\lambda_3,\lambda_4}{M_{p_\mathrm{t}}}{0,0,\lambda_1,\lambda_2}\nonumber\\
&= (2\pi)^2\frac{4E_{\rm cm}}{\sqrt{p_\mathrm{r} p'_\mathrm{r}}}\sum_{j=\max(\abs{\lambda},\abs{\lambda'})}^\infty\frac{2j+1}{4\pi} e^{i\lambda\phi}d^j_{\lambda\lambda'}(\theta)\mel*{\lambda_3,\lambda_4}{M^{j,m}_{p_\mathrm{t}}}{\lambda_1,\lambda_2}\nonumber\\
&=  \frac{4\pi E_{\rm cm}}{\sqrt{p_\mathrm{r} p'_\mathrm{r}}}e^{i\lambda\phi}\sum_{j=\max(\abs{\lambda},\abs{\lambda'})}^\infty(2j+1)d^j_{\lambda\lambda'}(\theta)\mel*{\lambda_3,\lambda_4}{M^{j}_{p_\mathrm{t}}}{\lambda_1,\lambda_2}\, ,
\end{align}
where $\lambda=\lambda_1-\lambda_2$, $\lambda'=\lambda_3-\lambda_4$. Due to the rotational symmetry, the amplitude is independent of the magnetic quantum number and only related to $j$, so in the last line we omitted the superscript $m$ of $M_{p_\mathrm{t}}^{j,m}$. Note that although $\mathbf{J}$ is a conserved quantity, $\mathbf{p}_\mathrm{r}$ and $\hat{\mathbf{p}}_\mathrm{r}$ are generally not conserved, and thus the total helicity is not conserved, hence there is no requirement for $\lambda=\lambda'$ in the summation of the above equation.

It is free to choose $xz$-plane as the $2\to 2$ scattering plane,
then the above equation can be simplified to
\begin{eqnarray}
M(\theta)=\frac{4\pi E_{\rm cm}}{\sqrt{p_\mathrm{r} p'_\mathrm{r}}}\sum_{j=\max(\abs{\lambda},\abs{\lambda'})}^\infty(2j+1)d^j_{\lambda\lambda'}(\theta)\mel*{\lambda_3,\lambda_4}{M^j_{p_\mathrm{t}}}{\lambda_1,\lambda_2}\,.
\end{eqnarray}
With the help of the orthogonality relation of $d$-functions, the inverse transformation can be obtained as
\begin{eqnarray}
\mel*{\lambda_3,\lambda_4}{M^j_{p_\mathrm{t}}}{\lambda_1,\lambda_2}=\frac{\sqrt{p_\mathrm{r} p'_\mathrm{r}}}{8\pi E_{\rm cm}}\int_0^\pi M(\theta) d^j_{\lambda\lambda'}(\theta)\sin\theta\dd{\theta}\,.
\end{eqnarray}

When the energy in the center-of-mass frame is much higher than the mass of each particle, we have $E_{\rm cm}/{\sqrt{p_\mathrm{r} p_\mathrm{r}'}}\approx2$. In this case, we obtain
\begin{eqnarray}
M(\theta)=8\pi\sum_{j=\max(\abs{\lambda},\abs{\lambda'})}^\infty(2j+1)d^j_{\lambda\lambda'}(\theta)\mel*{\lambda_3,\lambda_4}{M^j_{p_\mathrm{t}}}{\lambda_1,\lambda_2}\,.
\end{eqnarray}
For the scattering of zero helicity particles, we have $\lambda_1=\lambda_2=\lambda_3=\lambda_4=0$. 
Considering $d^j_{00}(\theta)=P_j(\cos\theta)$, we have
\begin{eqnarray}
M(\theta)=8\pi\sum_{j=0}^\infty(2j+1)\mel*{0,0}{M_{p_\mathrm{t}}^j}{0,0}P_j(\cos\theta)\,.
\end{eqnarray}

With the help of the orthogonality relation of the Legendre polynomials, the inverse transformation can be obtained as
\begin{eqnarray}
\mel*{0,0}{M_{p_\mathrm{t}}^j}{0,0}=\frac{1}{16\pi}\int_0^\pi M(\theta) P_j(\cos\theta)\sin\theta\dd{\theta}\,.
\end{eqnarray}
This result is what we need to use in the subsequent calculation of the upper bound of the dark Higgs mass.

\subsection{The case of identical particles\label{subsec:IdenticalParticals} }

The biggest difference between the identical particle state $\ket{\mathbf{p}_1,\eta_1;\mathbf{p}_2,\eta_2}$ and the non-identical particle state is their different normalization conditions:
\begin{align}
\braket{\mathbf{p}'_1,\eta_1';\mathbf{p}'_2,\eta_2'}{\mathbf{p}_1,\eta_1;\mathbf{p}_2,\eta_2} &= (2\pi)^6 4p_1^0p_2^0\delta^3(\mathbf{p}_1-\mathbf{p}'_1)\delta^3(\mathbf{p}_2-\mathbf{p}'_2) \delta_{\eta_1\eta'_1}\delta_{\eta_2\eta_2'}\nonumber\\
&\qquad \pm (2\pi)^6 4p_1^0p_2^0\delta^3(\mathbf{p}_1-\mathbf{p}'_2)\delta^3(\mathbf{p}_2-\mathbf{p}'_1) \delta_{\eta_1\eta'_2}\delta_{\eta_2\eta_1'}\,,
\end{align}
where the plus sign corresponds to bosons and the minus sign corresponds to fermions. The above equation can be proved by the commutation relation of the creation and annihilation operators. The sum or the difference of the two groups of $\delta$-functions in the above equation is the characteristic of identical particles. As can be seen from this point, unless $\eta_1=\eta_2=\eta_1'=\eta_2'$, only one or zero group of $\delta$-functions will remain in the final expression. This is easy to understand from a classical perspective. When $\eta_1\neq\eta_2$, the two particles can be distinguished by the internal quantum numbers, so they are not identical particles. For the sake of simplicity, we only consider the case where $\eta_1=\eta_2=\eta_1'=\eta_2'$ in the following, and omit the internal quantum numbers in them.

Applying Eq.~\eqref{eq:DeltaFunctionRelation} to the case of identical particles, we can get
\begin{eqnarray}
\braket{\mathbf{p}'_1,\mathbf{p}'_2}{\mathbf{p}_1,\mathbf{p}_2} = (2\pi)^6 \frac{4 E_{\rm cm}}{p_\mathrm{r}}\delta^4(p_\mathrm{t}-p_\mathrm{t}')\left(\delta(\Omega-\Omega')\pm\delta(\Omega-\Omega'')\right)\,,
\end{eqnarray}
where $\Omega''$ is the polar angle in the opposite direction of $\Omega'$ satisfying
\begin{eqnarray}
\delta(\Omega-\Omega'')=\delta(\cos\theta+\cos\theta')\delta(\phi-\phi'-\pi)\,.
\end{eqnarray}
Here $\delta(\phi-\phi'-\pi)$ may also be $\delta(\phi-\phi'+\pi)$, which is determined by the value of $\phi$. To avoid the confusion in it, we can rewrite $\delta(\Omega-\Omega')$ as
\begin{eqnarray}
\delta(\Omega-\Omega') = ie^{i\phi}\delta(\cos\theta-\cos\theta')\delta(e^{i\phi}-e^{i \phi'})\,.
\end{eqnarray}
In this way, for $\delta(\Omega-\Omega'')$ we have
\begin{eqnarray}
\delta(\Omega-\Omega'') = ie^{i\phi}\delta(\cos\theta+\cos\theta')\delta(e^{i\phi}+e^{i \phi'})\,.
\end{eqnarray}
With this representation, we can effectively avoid the problems caused by the periodicity of the polar angle.

Repeating the derivation in Sec.\ref{subsec:decomposition}, we will get
\begin{eqnarray}
\braket{\mathbf{p}'_\mathrm{r}}{\mathbf{p}_\mathrm{r}}=(2\pi)^2\frac{4E_{\rm cm}}{p_\mathrm{r}}\left(\delta(\Omega-\Omega')\pm\delta(\Omega-\Omega'')\right)\,.
\end{eqnarray}

$\ket{\theta,\phi}$ is still defined as
\begin{eqnarray}
\ket{\theta,\phi}=\frac{1}{2\pi}\sqrt{\frac{p_\mathrm{r}}{4E_{\rm cm}}}\ket{\mathbf{p}_{\mathrm{r}}}\,.
\end{eqnarray}
Therefore, we have
\begin{eqnarray}
\braket{\theta',\phi'}{\theta,\phi} = ie^{i \phi}\left(\delta(\cos\theta-\cos\theta')\delta(e^{i\phi}-e^{i \phi'})\pm \delta(\cos\theta+\cos\theta')\delta(e^{i\phi}+e^{i \phi'})\right)\,.
\end{eqnarray}
What needs to be known is that even in the case of identical particles, $\mel*{\mathbf{p}'_\mathrm{r}}{M_{p_\mathrm{t}}}{\mathbf{p}_\mathrm{r}}$ is still the scattering amplitude calculated by Feynman rules.

Partial wave expansion coefficient for identical particles case changes to
\begin{eqnarray}
C_j=\sqrt{1\pm(-1)^j}\sqrt{\frac{2j+1}{4\pi}}\,.
\end{eqnarray}
For identical bosons, the positive sign is taken in the above equation, so it is not zero only when $j$ is an even number. Thus we have 
\begin{eqnarray}
\ket{\theta,\phi}=\sqrt{2}\sum_{j=0,2,4,\cdots} \sqrt{\frac{2j+1}{4\pi}}\sum_{m=-j}^j  D^j_{m0}(\phi,\theta,0)\ket{j,m}\,.
\end{eqnarray}
Because identical particles satisfy $\lambda_1=\lambda_2$, so $\lambda=\lambda_1-\lambda_2=0$, the subscript 0 of the above rotation matrix reflects this fact. For identical fermions, we have
\begin{eqnarray}
\ket{\theta,\phi}=\sqrt{2}\sum_{j=1,3,5,\cdots} \sqrt{\frac{2j+1}{4\pi}}\sum_{m=-j}^j  D^j_{m0}(\phi,\theta,0)\ket{j,m}\,.
\end{eqnarray}
Whether for bosons or for fermions, the inverse transformation can be uniformly written as
\begin{eqnarray}
\ket{j,m}=\frac{1}{\sqrt{2}}\sqrt{\frac{2j+1}{4\pi}}\int \dd{\Omega}D^{j*}_{m0}(\theta,\phi,0)\ket{\theta,\phi}\,.
\end{eqnarray}
Compared with Eq.~\eqref{eq:inverse-partial-wave-exp}, it only has an additional $1/\sqrt{2}$ factor. With this feature, we can immediately know that the result obtained in the previous subsection should be rewritten as
\begin{equation}
\mel*{\lambda_3,\lambda_4}{M^j_{p_\mathrm{t}}}{\lambda_1,\lambda_2}=\left(\frac{1}{\sqrt{2}}\right)^n\frac{\sqrt{p_\mathrm{r} p'_\mathrm{r}}}{8\pi E_{\rm cm}}\int_0^\pi M(\theta) d^j_{\lambda\lambda'}(\theta)\sin\theta\dd{\theta}\,,
\end{equation}
where $n$ represents that there are $n$ pairs of identical particles in the incoming state and outgoing state.

\section{One-loop relation between renormalized parameters and physical parameters}\label{appB}

To obtain the pole masses of the dark photon and dark Higgs, we need to get the full propagators of the two fields under the $R_\xi$ gauge. We adopt the results in Ref.~\cite{Dudal:2019aew}. The transverse part of the dark photon full propagator is
\small 
\begin{eqnarray}
 \frac{1}{G_{A' A'}^{\perp}\left(p^2\right)}&=&p^2-m_{A'}^2
-\,\hbar \frac{2g'^2}{(4 \pi)^2} \int_0^1 \dd{x}\left\{K(m_{A'}^2, m_s^2)\left(1-\ln \frac{K(m_{A'}^2, m_s^2)}{\mu_r^2}\right) \right. \nonumber\\ 
&& \left. +m_s^2\left(1-\ln \frac{m_s^2}{\mu_r^2}\right)+\,\frac{m_{A'}^4}{m_s^2}\left(1-3 \ln \frac{m_{A'}^2}{\mu_r^2}\right)+2 m_{A'}^2 \ln \frac{K(m_{A'}^2, m_s^2)}{\mu_r^2}\right\}\nonumber\\
& \equiv & p^2-m_{A'}^2-\hbar\Pi_{A'A'}^{\perp}(p^2)\,,
\end{eqnarray}
\normalsize
where the function $K(m_1^2,m_2^2)$ implicitly depends on $x$ and $p^2$:
\begin{equation}
K(m_1^2,m_2^2)\equiv xm_1^2+(1-x)m_2^2-x(1-x)p^2\,.
\end{equation}
To confine the mass corrections to a fixed order, we cannot directly determine $m_{A',p}^2$ by setting
\begin{equation}\label{eq:OriginalPolemassEQ}
\Re((G_{A'A'}^\perp(m_{A'}^2))^{-1})=0\,.
\end{equation}
Since we only consider up to one-loop level, the $\mathcal{O}(\hbar^2)$ terms in Eq.~\eqref{eq:OriginalPolemassEQ} need to be ignored, resulting in
\begin{equation}
m_{A',p}^2-m_{A'}^2-\hbar\Re(\Pi_{A'A'}^{\perp}(m_{A',p}^2))=0\,.\label{eq:PoleMassEqForA}
\end{equation}
All the parameters of $\Pi_{A'A'}^{\perp}(m_{A',p}^2)$ in Eq.~\eqref{eq:PoleMassEqForA} can be directly taken as tree-level parameters or as physical parameters, since the $\mathcal{O}(\hbar^2)$ terms are ignored. Expressing the tree-level parameters as functions of physical parameters will facilitate our reparametrization of $\Delta V$. Hence we directly take the three parameters in $\Pi_{A'A'}^{\perp}(m_{A',p}^2)$ as corresponding physical parameters, obtaining
\small
\begin{equation}
\begin{aligned}
m_{A'}^2 &= m_{A',p}^2-\hbar\Re(\Pi_{A'A'}^{\perp}(m_{A',p}^2))\\
&= m_{A',p}^2+\frac{\hbar g'^2}{8\pi^2}\Re\Bigg[\frac{31 m_{A',p}^2}{9}-\frac{5 m_{s,p}^2}{2}-\frac{m_{A',p}^4}{m_{s,p}^2}+\frac{m_{s,p}^4}{6 m_{A',p}^2}+m_{s,p} \sqrt{4 m_{A',p}^2-m_{s,p}^2}\,\\
&\qquad \times\left(\frac{2 m_{s,p}^2}{3 m_{A',p}^2}-\frac{m_{s,p}^4}{6 m_{A',p}^4}-2\right) \Bigg(\arctan\frac{m_{s,p}}{\sqrt{4 m_{A',p}^2-m_{s,p}^2}}+\arctan\frac{2 m_{A',p}^2-m_{s,p}^2}{\sqrt{4 m_{A',p}^2 m_{s,p}^2-m_{s,p}^4}}\Bigg)\\
&\qquad +\,\left(\frac{m_{s,p}^6}{12 m_{A',p}^4}-\frac{m_{s,p}^4}{2 m_{A',p}^2}\right) \ln\frac{m_{A',p}^2}{m_{s,p}^2}+\left(\frac{3 m_{A',p}^4}{m_{s,p}^2}-\frac{5 m_{A',p}^2}{3}+\frac{3 m_{s,p}^2}{2}\right) \ln\frac{m_{A',p}^2}{\mu ^2}\Bigg]\,.
\end{aligned}
\end{equation}
\normalsize 

On the other hand, the full propagator of the dark Higgs is
\small
\begin{align}
\frac{1}{G_{ss}(p^2)} &= p^2-m_s^2+\frac{\hbar}{(4 \pi)^2} \int_0^1 \dd{x}\left\{g'^ { 2 } \left[p^2\left(\ln \frac{m_{A'}^2}{\mu_r^2}+2 \ln \frac{K\left(m_{A'}^2, m_{A'}^2\right)}{\mu_r^2}-1\right)\right.\right. \notag\\
& \qquad\left.-\,\frac{(p^2)^2}{2 m_{A'}^2} \ln \frac{K\left(m_{A'}^2, m_{A'}^2\right)}{\mu_r^2}-6 m_{A'}^2\left(1-\ln \frac{m_{A'}^2}{\mu_r^2}+\ln \frac{K\left(m_{A'}^2, m_{A'}^2\right)}{\mu_r^2}\right)\right] \notag\\
& \qquad+\,\frac{\lambda}{4} m_s^2\left(-6+6 \ln \frac{m_s^2}{\mu_r^2}-9 \ln \frac{K\left(m_s^2, m_s^2\right)}{\mu_r^2}\right) \notag\\
& \qquad\left.-\,\left[\xi\left(\frac{\lambda}{2} m_{A'}^2-g'^2 p^2\right)\left(1-\ln \frac{\xi m_{A'}^2}{\mu_r^2}\right)-\left(g'^2 \frac{(p^2)^2}{2 m_{A'}^2}-\frac{\lambda}{4}m_s^2\right) \ln \frac{K\left(\xi m_{A'}^2, \xi m_{A'}^2\right)}{\mu_r^2}\right]\right\}\notag\\
&\equiv p^2-m_s^2-\hbar\Pi_{ss}(p^2)\,,
\end{align}
\normalsize 
where $\lambda$ needs to be replaced with $2 g'^2 m_s^2/m_{A'}^2$. Thus, we can obtain
\begin{equation}
\begin{aligned}
m_s^2 &= m_{s,p}^2-\hbar\Re(\Pi_{ss}(m_{s,p}^2))\\
&= m_{s,p}^2-\hbar \frac{g_p'^2}{32\pi^2}\frac{m_{s,p}^2}{m_{A',p}^2}\Re\Bigg[10m_{A',p}^2-12\frac{m_{A',p}^4}{m_{s,p}^2}+\left(3\sqrt{3}\pi-14\right)m_{s,p}^2\\
&\qquad +\, 2m_{s,p}\sqrt{4 m_{A',p}^2-m_{s,p}^2}\left(1-4\frac{m_{A',p}^2}{m_{s,p}^2}+12\frac{m_{A',p}^4}{m_{s,p}^4}\right)\arctan\frac{m_{s,p}}{\sqrt{4 m_{A',p}^2-m_{s,p}^2}}\\
&\qquad\left.+\,(m_{s,p}^2-6m_{A',p}^2)\ln\frac{m_{A',p}^2}{\mu_r^2}+3 m_{s,p}^2\ln\frac{m_{s,p}^2}{\mu_r^2}\right]\,.
\end{aligned}
\end{equation}
As expected, although $G_{ss}(p^2)$ is gauge-dependent, the pole mass is actually gauge independent.

\begin{figure}[ht]
\centering
\includegraphics[width=0.7\textwidth]{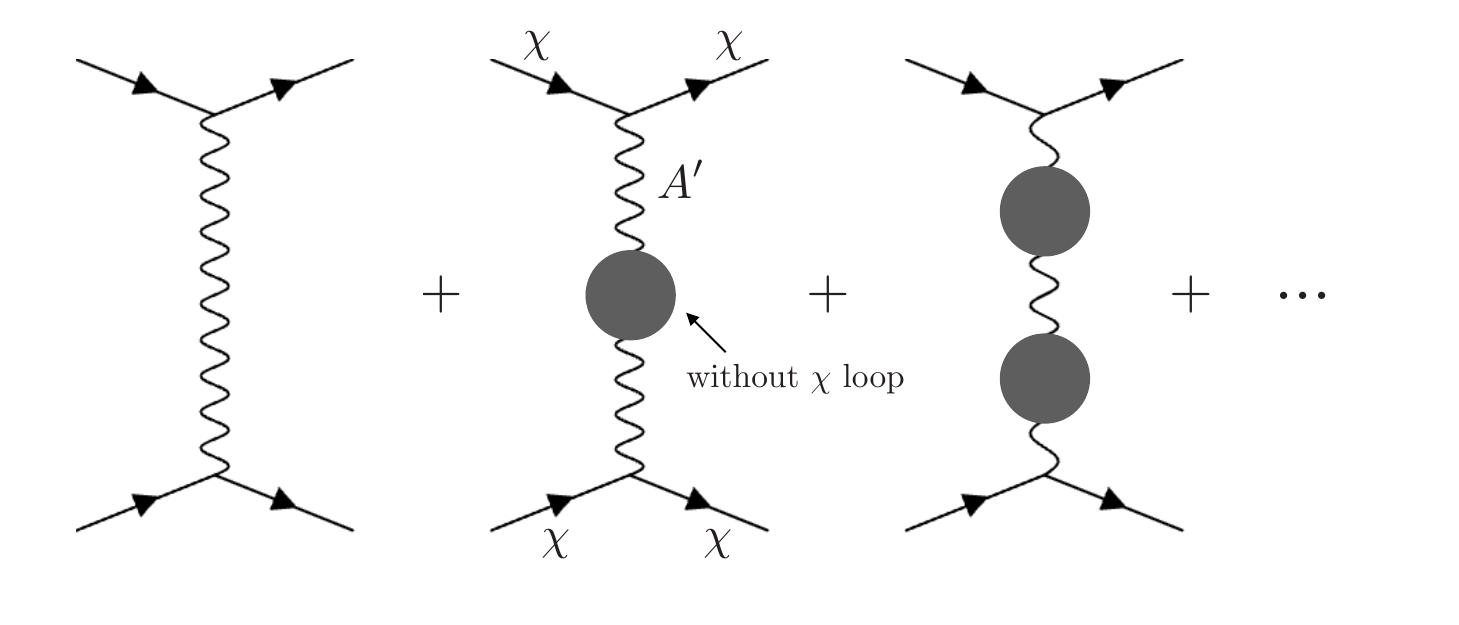}
\caption{Feynman diagrams considered for the definition of $g'_p$.}
\label{fig:FeynmannGraphics4gp}
\end{figure}

Next, we need to express $g'$ as a function of physical parameters, and the result depends on how we define $g_p'$. Since our model originates from the study of DM, we can define $g_p'$ through the scattering of DM particles. 
We consider the DM pair elastic scattering process $\chi+\chi\to\chi+\chi$ with the momentum transfer $q$ much smaller than DM mass $m_\chi$, i.e. $q^2 \ll m_\chi^2$. 
This situation is generally true when we study DM scattering in halos.
Because the DM mass $m_\chi$ is much larger than all other energy or masses in $\chi+\chi\to\chi+\chi$ process, the dominate contribution is actually the t-channel process. Other processes like u-channel or box diagram can all be ignored. 
Furthermore, for loop corrections, we only need to consider vacuum polarization of $A'$ without DM $\chi$ loop.   
Fig.~\ref{fig:FeynmannGraphics4gp} is an illustration of Feynman diagrams we considered.  

At one-loop level, the spin-averaged amplitude square is proportional to coupling and full-propagator: 
\begin{equation}
\sum_{\text{spins}}\abs{\mathcal{M}_{\text{1L}}\left(\chi(p_1)+\chi(p_2)\to \chi(p_1')+\chi(p_2')\right)}^2= \frac{g'^4}{ q^2 - m^2_{A'} -\hbar\Pi_{A'A'}^\perp (q^2) } \times F \, ,
\end{equation}
with $q=p_1 - p'_1$ and 
\begin{equation}
F\equiv \tr[(\slashed{p}_1'+m_\chi)\gamma^\mu(\slashed{p}_1+m_\chi)\gamma^\nu]\tr[(\slashed{p}_2'+m_\chi)\gamma_\mu(\slashed{p}_2+m_\chi)\gamma_\nu]\,.
\end{equation}

Then the physical coupling $g'_p$ can be defined as the effective coupling when momentum transfer $q^2 = - m^2_{A',p}$:
\begin{equation}
\left. \sum_{\text{spins}}\abs{\mathcal{M}\left(\chi(p_1)+\chi(p_2)\to \chi(p_1')+\chi(p_2')\right)}^2 \right|_{q^2 = - m^2_{A',p}}= \frac{{g'_p}^4}{ (- m^2_{A',p}) - m^2_{A',p}  } \times F 
\end{equation}
That is to say, low velocity $\chi\chi$ scattering is carried out through Yukawa potential $V(r) = \frac{{g'_p}^2}{4\pi}\frac{1}{r} e^{- m_{A',p} r}$ at length scale around $1/m_{A',p}$.

Then we find the one-loop order relation between $g'$ and $g'_p$: 
\begin{equation}\label{eq:AmplitudeAtDefPoint}
\frac{g_p'^4F}{\abs{(-m_{A',p}^2)-m_{A',p}^2}^2}=\frac{g'^4F}{\abs{(-m_{A',p}^2)-m_{A'}^2-\hbar\Pi_{A'A'}^\perp(-m_{A',p}^2)}^2}\, ,
\end{equation}
and $g'^2$ can be expressed by physical parameters: 
\begin{equation}
g'^2=g'^2_p\left(1+\frac{\hbar}{2 m_{A',p}^2}\left(\Re\Pi_{A'A'}^\perp(-m_{A',p}^2)-\Re\Pi_{A'A'}^\perp(m_{A',p}^2)\right)\right)+\mathcal{O}(\hbar^2)\,.
\end{equation}


\bibliography{ref}

\end{document}